\documentclass[useAMS,usenatbib]{mnras}
\usepackage{natbib}
\usepackage{amsmath}
\usepackage{amssymb}
\usepackage{graphicx}
\usepackage{multirow}
\usepackage{url}
\usepackage{enumitem}
\usepackage[table]{xcolor}

\newcommand{\vspBottom}{\vspace{-0.15 in}}

\title[Multiphase AGN winds]{Unravelling the physics of multiphase AGN winds through emission line tracers}
\author[A. J. Richings, C.-A. Faucher-Gigu\`{e}re and J. Stern]{Alexander J. Richings$^{1}$\thanks{Email: alexander.j.richings@durham.ac.uk}, Claude-Andr\'{e} Faucher-Gigu\`{e}re$^{2}$ and Jonathan Stern$^{2, 3}$\\
$^{1}$Institute for Computational Cosmology, Department of Physics, Durham University, South Road, Durham, DH1 3LE, United Kingdom \\
$^{2}$Center for Interdisciplinary Exploration and Research in Astrophysics (CIERA) and Department of Physics and Astronomy,\\ 
  Northwestern University, 1800 Sherman Ave, Evanston, IL 60201, USA\\
$^{3}$School of Physics and Astronomy, Tel Aviv University, Tel Aviv 69978, Israel}

\begin{document}

\date{Accepted 2021 February 20. Received 2021 February 19; in original form 2020 December 11.}

\pagerange{\pageref{firstpage}--\pageref{lastpage}} \pubyear{2021}

\maketitle

\label{firstpage}

\begin{abstract} 
  Observations of emission lines in Active Galactic Nuclei (AGN) often find fast ($\sim$1000~km~s$^{-1}$) outflows extending to kiloparsec scales, seen in ionised, neutral atomic and molecular gas. In this work we present radiative transfer calculations of emission lines in hydrodynamic simulations of AGN outflows driven by a hot wind bubble, including non-equilibrium chemistry, to explore how these lines trace the physical properties of the multiphase outflow. We find that the hot bubble compresses the line-emitting gas, resulting in higher pressures than in the ambient ISM or that would be produced by the AGN radiation pressure. This implies that observed emission line ratios such as [O\textsc{iv}]$_{25 \, \rm{\mu m}}$ / [Ne\textsc{ii}]$_{12 \, \rm{\mu m}}$, [Ne\textsc{v}]$_{14 \, \rm{\mu m}}$ / [Ne\textsc{ii}]$_{12 \, \rm{\mu m}}$ and [N\textsc{iii}]$_{57 \, \rm{\mu m}}$ / [N\textsc{ii}]$_{122 \, \rm{\mu m}}$ constrain the presence of the bubble and hence the outflow driving mechanism. However, the line-emitting gas is under-pressurised compared to the hot bubble itself, and much of the line emission arises from gas that is out of pressure, thermal and/or chemical equilibrium. Our results thus suggest that assuming equilibrium conditions, as commonly done in AGN line emission models, is not justified if a hot wind bubble is present. We also find that $\gtrsim$50 per cent of the mass outflow rate, momentum flux and kinetic energy flux of the outflow are traced by lines such as [N\textsc{ii}]$_{122 \, \rm{\mu m}}$ and [Ne\textsc{iii}]$_{15 \, \rm{\mu m}}$ (produced in the 10$^{4} \, \rm{K}$ phase) and [C\textsc{ii}]$_{158 \, \rm{\mu m}}$ (produced in the transition from 10$^{4} \, \rm{K}$ to 100 K). 
\end{abstract}


\section{Introduction}\label{intro_sect} 

Galaxies that host active galactic nuclei (AGN) have been observed to contain fast ($\sim$1000~km~s$^{-1}$) outflows of gas on kiloparsec scales \citep[e.g.][]{cecil88, veilleux03}, likely driven by the input of energy and momentum from accretion onto the central supermassive black hole \citep[e.g.][]{feruglio10, cicone14}.

Given their high velocities, one might expect the outflowing material to be almost exclusively hot ($\gtrsim$10$^{7} \, \rm{K}$). However, these outflows have been observed in a wide range of emission and absorption lines, spanning molecular, neutral atomic and ionised gas phases \citep{fischer10, rupke11, greene12, harrison12, harrison14, aalto15, fiore17, fluetsch19, feruglio20, lutz20}.

Several possible mechanisms for the formation of the cool phase in outflows have been considered (applicable to both AGN- and star formation-driven winds), including the entrainment of cold gas clouds in a hot outflow \citep{scannapieco15, gaspari17, schneider17, gronke20}, the direct acceleration of cold gas by cosmic rays or radiation pressure \citep{booth13, costa18, hopkins20}, or in-situ cooling of the hot outflowing material \citep{wang95, silich03, zubovas14, costa15, thompson16}. While the true origin of the cool phase in these powerful outflows remains uncertain, the observational evidence for such a multiphase structure is nonetheless overwhelming (see also \citealt{veilleux20} for a recent review). 

AGN-driven galactic outflows are important as they are likely to play a vital role in shaping the formation and evolution of galaxies. The energy and momentum injected by black hole winds can regulate the growth of the black holes and the stellar component of their host galaxies, giving rise to the observed scaling relations between the two \citep{silk98, king03, murray05, zubovas12, torrey20}. Energetic feedback from AGN can quench star formation in the most massive galaxies \citep{binney95, dubois13, bower17}, and is required by modern cosmological models of galaxy formation to reproduce observed galaxy populations in terms of their stellar masses (\mbox{\citealt{bower12}}; \mbox{\citealt{crain15}}; \mbox{\citealt{tremmel17}}; \mbox{\citealt{weinberger18}}; \mbox{\citealt{dave19}}), the colours of massive elliptical galaxies \mbox{(\citealt{springel05};} \citealt{feldmann17}; \citealt{trayford16}), and the stellar mass densities of massive galaxies \citep{choi18, wellons20, parsotan20}. Outflows can also enrich the circumgalactic medium (CGM) through the transport of metals from the galaxy \citep{hummels13, ford13, muratov17, tumlinson17, hafen19}, and can deplete CGM gas fractions by carrying baryons beyond the virial radius \citep{davies19, oppenheimer20}. 

To understand how such outflows can influence the surrounding environment, we need to measure their energetics (e.g. mass outflow rates, momentum fluxes, and energy fluxes), to determine whether they contain sufficient energy and momentum to have a significant impact on their host galaxy. This requires us to connect the emission and absorption line tracers in which the outflows have been observed to the physical properties of their constituent gas components.

Many studies have successfully employed photoionisation models to reproduce the emission line properties of AGN. \citet{dopita02} and \citet{groves04a} presented photoionisation models of dusty clouds dominated by radiation pressure in the Narrow Line Region (NLR) of AGN. The pressure and density structure of the clouds were determined by assuming that they are in hydrostatic balance with the radiation pressure exerted by the AGN, also known as Radiation Pressure Confinement (RPC; see also \citealt{draine11} and \citealt{yeh12} for the same effect in ionised gas around star forming regions). They then used the \textsc{mappings iii} photoionisation and shock code to calculate the intensities of emission lines on a grid of density, metallicity, ionisation parameter and the power-law index of the ionising spectrum. Observed line ratio diagnostics can then be compared to these models to deduce the physical properties of the line-emitting clouds \citep{groves04b}. 

\citet{stern14a} used the photoionisation code \textsc{cloudy} to model the emission from RPC clouds spanning a large range of distances from the nucleus. They showed that RPC can explain the observed optical emission line ratios and the overlap of extended X-ray and optical line emission in nearby Seyferts. \citet{bianchi19} later showed that these types of models can explain the emission measure distribution of X-ray emission lines. RPC models also explain observed Broad Line Region (BLR) emission line ratios over a range of $10^{8}$ in AGN luminosity \citep{baskin14} and the broad ionisation distribution in AGN outflows \citep{stern14b}. \citet{stern16} used these RPC models and observed emission line ratios in luminous quasars to constrain the relative importance of hot gas and radiation pressure on the dynamics of quasar outflows. They found no evidence for the compression of emission line gas expected in the presence of a hot bubble on any scale, and argued that a dynamically important hot bubble can be ruled out on spatial scales below 40 pc.

Such photoionisation models have had success in reproducing many of the observed emission line properties of AGN. However, they assume pressure, thermal and/or chemical equilibrium in the line-emitting gas. These assumptions need to be tested further. 

In \citet{richings18a} (hereafter RFG18), we ran a suite of hydro-chemical simulations of AGN winds to explore the origin of molecular outflows. We demonstrated that observed molecular outflows in AGN host galaxies can be produced by the in-situ formation of new molecules within the outflowing material. In that work, we focussed on the molecular phase. However, the chemical modelling in these simulations also includes the chemistry of the atomic and ionised phases. This enables us to make direct predictions for the emission lines from all phases of the outflow. Most previous hydrodynamic simulations of AGN outflows do not directly model the line emission, which is important as emission lines contain most of the observational constraints on the models. Our simulation suite therefore presents a unique tool with which to study the connection between the physical properties and energetics of multiphase AGN outflows and the observable emission line tracers, and to test the underlying assumptions that enter into alternative photoionisation models for AGN emission. 

In this paper, we compute emission lines from the ionised, neutral atomic and molecular phases of the AGN outflows in the RFG18 simulations, which we use to explore the physical properties of the line-emitting gas. The remainder of this paper is organised as follows. In Section~\ref{methods_sect} we describe the simulations and the radiative transfer calculations used to model the line emission. In Section~\ref{emission_sect} we present predictions for the line emission from the simulations (Section~\ref{predict_sect}), investigate the physical properties of the line-emitting gas (Section~\ref{properties_sect}), and compare our model predictions to observations (Section~\ref{comparison_sect}). We explore how the emission line ratios computed from our simulations can be used to constrain the driving mechanisms of AGN outflows in Section~\ref{drive_sect}, and in Section~\ref{energetics_sect} we study the outflow energetics of the different gas phases traced by the various emission lines. Finally, we summarise our results in Section~\ref{conclusions_sect}. 

\vspBottom

\section{Methods}\label{methods_sect}  

\subsection{Simulations}\label{sims_sect}

In RFG18 we presented a series of hydro-chemical simulations of AGN-driven galactic outflows. These simulations model an initially uniform ambient medium, into which we inject an isotropic AGN wind by spawning gas particles in the central parsec with an outward velocity of $30 \, 000 \, \rm{km} \, \rm{s}^{-1}$ and a momentum injection rate determined by the AGN luminosity, $L_{\rm{AGN}}$. Each simulation follows the interaction of this wind with the ambient medium over $1 \, \rm{Myr}$, which corresponds to the typical flow times ($r / v$) of kiloparsec-scale outflows that have been observed in luminous quasars \citep[e.g.][]{gonzalezalfonso17}. These simulations use the gravity+hydrodynamics code \textsc{gizmo}, with the Meshless Finite Mass (MFM) hydro solver \citep{hopkins15} and a fiducial resolution of $30 \, \rm{M}_{\odot}$ per gas particle. 

We model the time-dependent chemistry of the gas using the \textsc{chimes} non-equilibrium chemistry and cooling module\footnote{\url{https://richings.bitbucket.io/chimes/home.html}} \citep{richings14a, richings14b}, which follows the evolution of 157 ions and molecules that dominate the cooling rate from cold ($\sim$10 K), molecular gas to hot ($>$10$^{9}$ K), highly ionised plasmas. The \textsc{chimes} chemical network contains various collisional reactions, including collisional ionisation, recombination, charge transfer reactions, and reactions on the surface of dust grains such as the formation of molecular hydrogen (assuming a constant dust-to-metals ratio), along with photoionisation and photodissociation reactions.

For the photochemistry, we use the average quasar spectrum from \citet{sazonov04} normalised according to the bolometric AGN luminosity and the distance of the gas particle from the black hole. This is an average between an obscured and an unobscured spectrum, and is characteristic of a typical quasar. We chose this spectrum as our simulations focus on outflowing material at kiloparsec scales, but do not explicitly model the small-scale structures around the AGN that are likely to contribute to the obscuration of the AGN radiation. By using an average spectrum in this way, our aim was to capture the partial obscuration at small scales. This spectrum differs from that used in some other models of AGN emission lines. For example, \citet{stern16} use a power-law spectrum representative of an unobscured quasar, with a fiducial extreme UV (EUV) slope of $-1.6$ (along with variations in this slope). In particular, the UV to X-ray ratio is a factor $\approx$2 lower in the average \citet{sazonov04} spectrum than the fiducial power-law spectrum of \citet{stern16}. We will consider the effects of the choice of spectrum below. 

Self shielding is included using a Sobolev-like shielding length approximation based on the density gradient (see equation~5 of RFG18). The densities of individual species are multiplied by this shielding length to obtain their column densities, which are then used to attenuate the photoionisation, photodissociation and photoheating rates. 

Further details of our simulations can be found in RFG18. For this paper, we focus on the low-luminosity (nH10\_L45\_Z1) and fiducial (nH10\_L46\_Z1) runs. Both use a uniform ambient medium with an initial hydrogen density $n_{\rm{H}} = 10 \, \rm{cm}^{-3}$ and solar metallicity\footnote{Throughout this paper, we use the solar abundances listed in table~1 of \citet{wiersma09}, for which the solar metallicity is $\rm{Z}_{\odot} = 0.0129$.}, but they differ in the bolometric AGN luminosity, which is $10^{45}$ and $10^{46} \, \rm{erg} \, \rm{s}^{-1}$, respectively. For brevity, we will refer to these two runs as L45 and L46 for the remainder of this paper. The low-density run ($n_{\rm{H}} = 1 \, \rm{cm}^{-3}$) from RFG18 did not cool and form a multiphase wind before the end of the simulation after $1 \, \rm{Myr}$, while the low-metallicity run ($Z = 0.1 Z_{\odot}$) strongly under-predicts the molecular outflow rates compared to observations. 

\vspBottom

\subsection{Radiative transfer calculations}\label{rt_sect} 

To create maps of the emission lines from our simulations, we post-process the simulation snapshots using version 0.40 of the publicly available Monte Carlo radiative transfer code \textsc{radmc-3d}\footnote{\url{http://www.ita.uni-heidelberg.de/~dullemond/software/radmc-3d/}} \citep{dullemond12}, using the non-equilibrium ion and molecule abundances that were calculated during the simulations with the \textsc{chimes} chemistry module. While the full 3D spherical outflow is included in the simulations, only one octant of the simulation volume uses the highest resolution level, with the remainder of the volume using 8$\times$ lower mass resolution. We therefore only use the high-resolution octant to produce the emission line maps. This octant is mirrored in the line of sight direction, to capture both the receding and approaching sides of the outflow, as viewed by the observer. The resulting emission maps thus cover one quadrant of the outflow. 

As \textsc{radmc-3d} is a grid-based code, we first project the gas particles from the simulations on to an Adaptive Mesh Refinement (AMR) grid, which is refined such that no cell contains more than 8 particles. The particles are smoothed using a cubic spline kernel with a smoothing length enclosing 32 neighbours, as used in the MFM hydro solver for these simulations. When projecting the gas temperatures and velocities on to the grid, we weight the contribution from each particle by the given ion or molecule abundance, to avoid unphysical effects from mixing particles with very different properties in the same cell (see the discussion in section 5 of RFG18). 

We calculate the emissivities of H$\alpha$ and H$\beta$ including both recombination of H\textsc{ii} and collisional excitation of H\textsc{i}, using the cascade matrix formalism described in \citet{raga15}, with the atomic data and fits for the collision strengths and recombination coefficients that they present in their Appendix~A. For all other ions and molecules, the level populations are computed by \textsc{radmc-3d} using an approximate non-LTE approach based on the Local Velocity Gradient (LVG) method. We utilised atomic data (energy levels, transition probabilities and collisional excitation rates) from version 7.1 of the \textsc{chianti} database \footnote{\url{https://www.chiantidatabase.org/chianti.html}} \citep{dere97, landi13}, supplemented with additional collisional excitation rate data from the \textsc{lamda} database\footnote{\url{https://home.strw.leidenuniv.nl/~moldata/}} \citep{schoier05} for C\textsc{i}, C\textsc{ii} and O\textsc{i}.

Dust absorption has little to no effect on the infrared emission lines, but those at optical and UV wavelengths are strongly affected by dust. The idealised setup of these simulations, in which the outflow is embedded within a dense ambient interstellar medium (ISM), is representative of an AGN in the buried QSO phase, before the outflow has broken out of the disk of its host galaxy. In observations of buried QSOs, the outflows are very difficult to detect in optical and UV lines due to the strong dust absorption from the surrounding ISM. Even in systems where the outflows have broken out of the dense medium, they are typically only seen in the blue wing of the emission line (i.e. from the near side of the outflow, moving towards the observer). 

When we include the effects of dust in our radiative transfer calculations, we also find that the optical and UV lines are strongly suppressed, with the outflow only seen in the blue wings. However, in our idealised setup the host galaxy is represented by a box of initially uniform density gas up to 2.4~kpc across, which does not include the turbulent structures we would expect to see in the ISM nor the geometry of a galactic disc. We therefore cannot model the realistic dust attenuation from the host galaxy in this setup. Including dust absorption from the ambient medium also limits how much of the optical and UV emission we can study from the simulations. 

For the radiative transfer calculations in this paper, we therefore include dust grains only in the outflow. For gas with a radial velocity $\leq$0~km~s$^{-1}$ (i.e. the ambient ISM, which is slowly inflowing by the end of the simulation due to the gravitational potential of the host galaxy), we set the dust density to zero. We also disable dust when the gas temperature is above $10^{6} \, \rm{K}$, as grains will be rapidly destroyed by sputtering in this regime \citep[e.g.][]{tsai95}. This approach has little impact on the infrared lines. However, we stress that the resulting optical and UV line luminosities do not include the effects of dust attenuation from the host galaxy. When comparing these lines to observations, it is therefore important that we only consider line ratios at similar wavelengths, for which the strength of any additional dust absorption on each line will be similar, thus leaving the ratios unaffected. By excluding dust from the ambient medium in this way, we can focus on the nature of the gas that produces each emission line.

For the dust in the outflow, we use a mixture of graphite and silicate grains with dust-to-gas ratios of $2.4 \times 10^{-3}$ and $4.0 \times 10^{-3}$, respectively, at solar metallicity. We include dust absorption, scattering and thermal emission. To obtain the continuum emission, we repeat each \textsc{radmc-3d} calculation a second time with the emission lines disabled. The resulting continuum emission is then subtracted from the total emission. 

\vspBottom

\section{Emission line predictions}\label{emission_sect}

\subsection{Spectra and images}\label{predict_sect}

\begin{figure}
\centering
\mbox{
	\includegraphics[width=84mm]{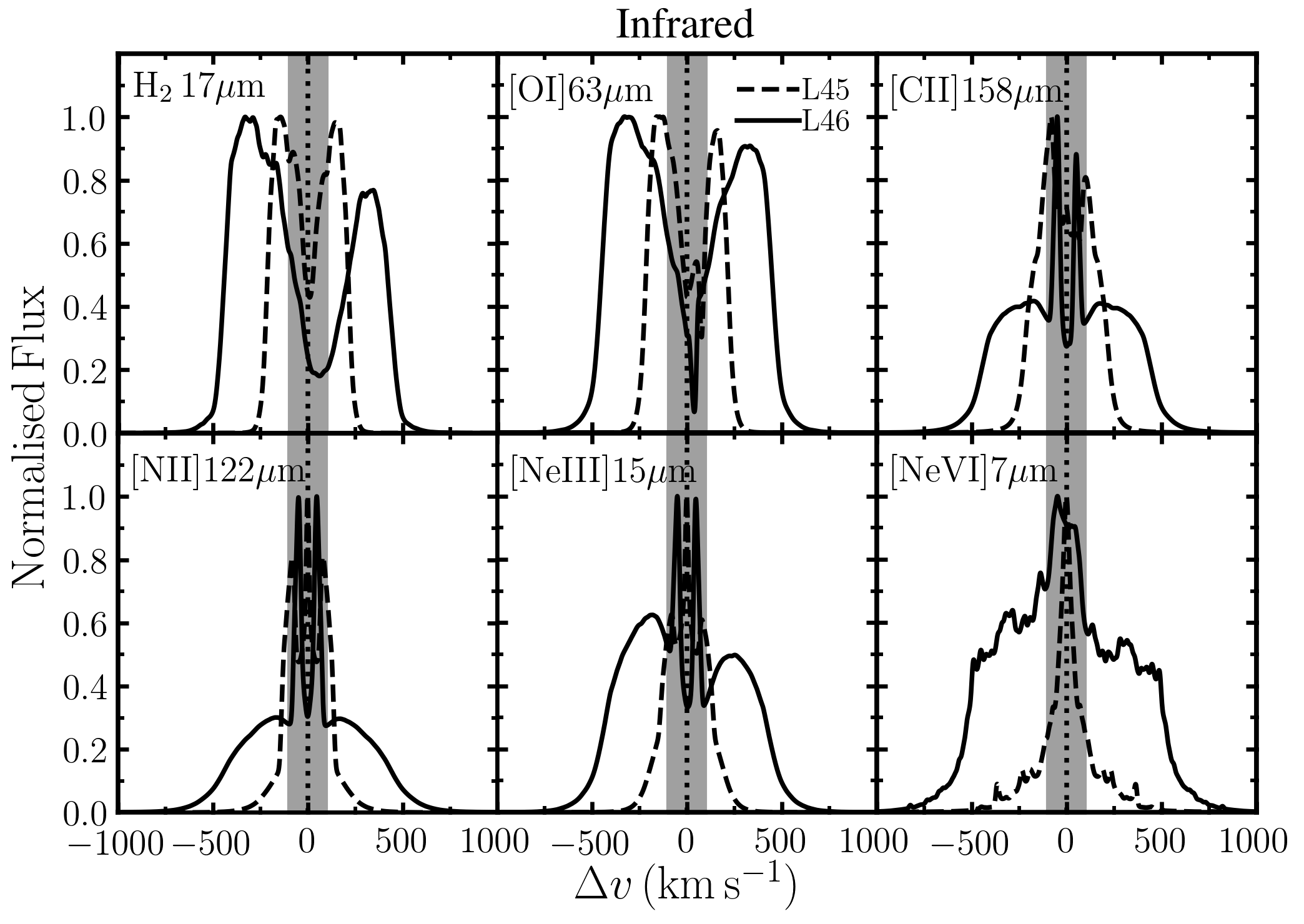}}
\mbox{
	\includegraphics[width=84mm]{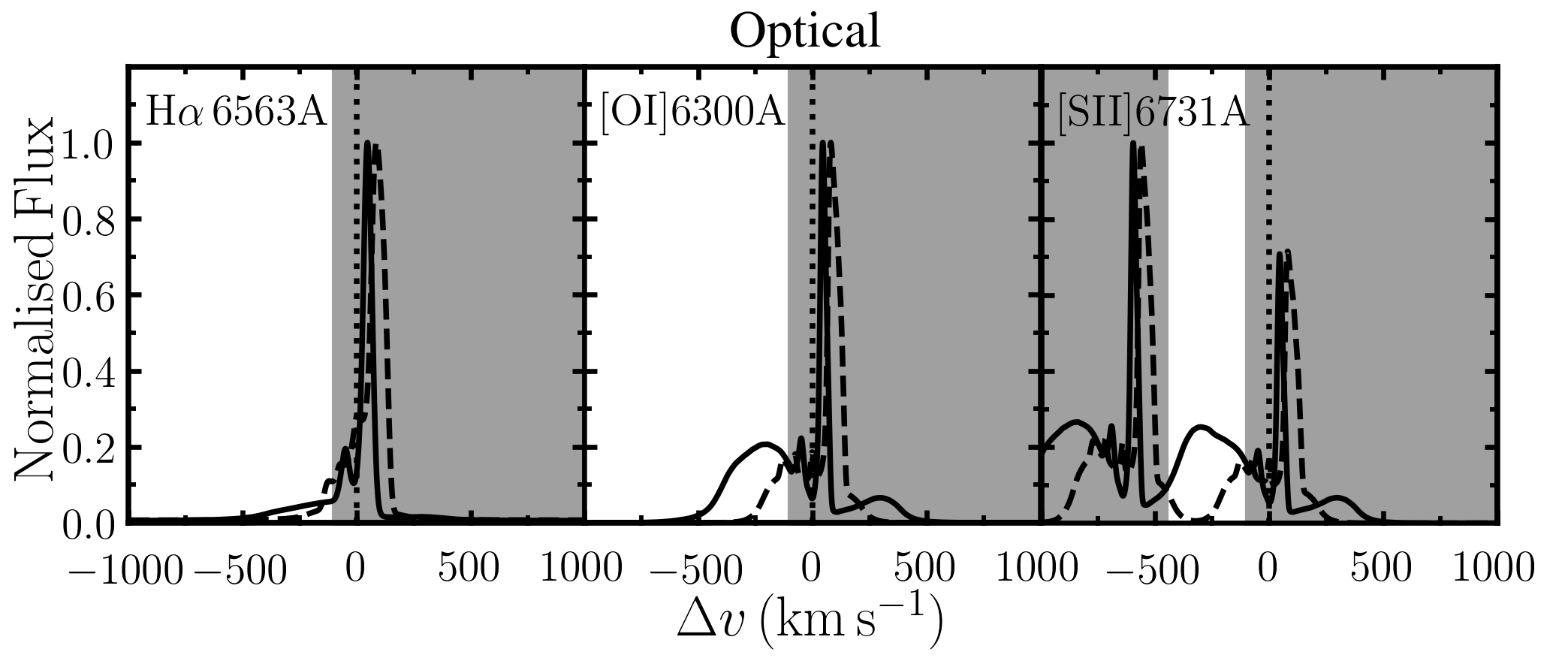}}
\vspace{-0.15 in}
\caption{Continuum-subtracted spectra of six infrared emission lines (top two rows) and three optical emission lines (bottom row) commonly observed in AGN, from the simulations L45 (dashed curves) and L46 (solid curves), normalised to the maximum flux in each spectrum. These represent a subset of the lines that we study in this paper; the spectra from the full sample of 24 lines can be found in Appendix~\ref{full_set_sect}. The vertical dotted line in each panel indicates the line centre. The broad component of these lines arises from the outflowing shell, with velocities extending up to $\approx$300 and $\approx$700~km~s$^{-1}$ in L45 and L46, respectively. However, central velocities at $|\Delta v|$$\leq$$100 \, \rm{km} \, \rm{s}^{-1}$ are typically dominated by narrow components from the ambient ISM. The grey shaded bands indicate velocity ranges that are excluded for the remainder of our analysis, to focus on the outflow component. 
\vspace{-0.15 in}} 
\label{spectra}
\end{figure}

Fig.~\ref{spectra} shows the spectra of six infrared emission lines (top two rows) and three optical lines (bottom row) that are commonly observed in AGN host galaxies. These represent a subset of the lines that we study in this paper; the full sample of 24 spectra can be found in Appendix~\ref{full_set_sect}. The dashed and solid curves in each panel are from the simulations L45 and L46, respectively. Each spectrum is only integrated over the spatial extent of the outflow, out to a radius of 0.57 kpc (L45) and 1.04 kpc (L46). The spatial extent in each case is determined by the radius of the outflowing shell after 1 Myr, when the simulation ends. Thus the L46 simulation reaches a larger spatial extent due to the higher outflow velocity in this case. We focus our analysis at 1 Myr because this corresponds to the typical flow times, $t_{\rm{flow}} = r / v$, of outflows observed in luminous AGN \citep[e.g.][]{gonzalezalfonso17}. The fluxes are normalised to the maximum flux in each spectrum.

All of these emission lines show a broad component from the outflowing shell, with velocities up to $\approx$300 and $\approx$700~km~s$^{-1}$ in L45 and L46, respectively. However, several of these lines show a strong narrow component arising from the ambient medium, which is seen either in emission (e.g. [C\textsc{ii}]$_{158 \, \rm{\mu m}}$) or absorption (e.g. [O\textsc{i}]$_{63 \, \rm{\mu m}}$). For the infrared lines, we therefore only consider emission in the wings of each line for the rest of our analysis, with velocities $|\Delta v|$$>$$100 \, \rm{km} \, \rm{s}^{-1}$, to focus on the outflow component.

In the optical lines (and UV lines; not shown), the outflow is seen more strongly in the blue wing (approaching) than the red wing (receding) due to strong dust absorption at these wavelengths, as noted in Section~\ref{rt_sect}. We again stress that we only include dust grains in the outflow for these radiative transfer calculations. This effect would be even more dramatic when we include the additional dust absorption from the ambient ISM in the host galaxy. For our further analysis of the optical and UV lines, we therefore only consider the blue wing, with velocities $\Delta v$$<$$-100 \, \rm{km} \, \rm{s}^{-1}$. As the [S\textsc{ii}] lines are a doublet, we also need to exclude velocities $<$$-450 \, \rm{km} \, \rm{s}^{-1}$ to avoid contamination of the 6731\AA \, line by the 6716\AA \, line. We apply this additional velocity cut to both of the sulphur lines. The velocity ranges that are excluded from the further analysis of each line are highlighted by the grey shaded bands in Fig.~\ref{spectra}. 

\begin{figure}
\centering
\mbox{
	\includegraphics[width=84mm]{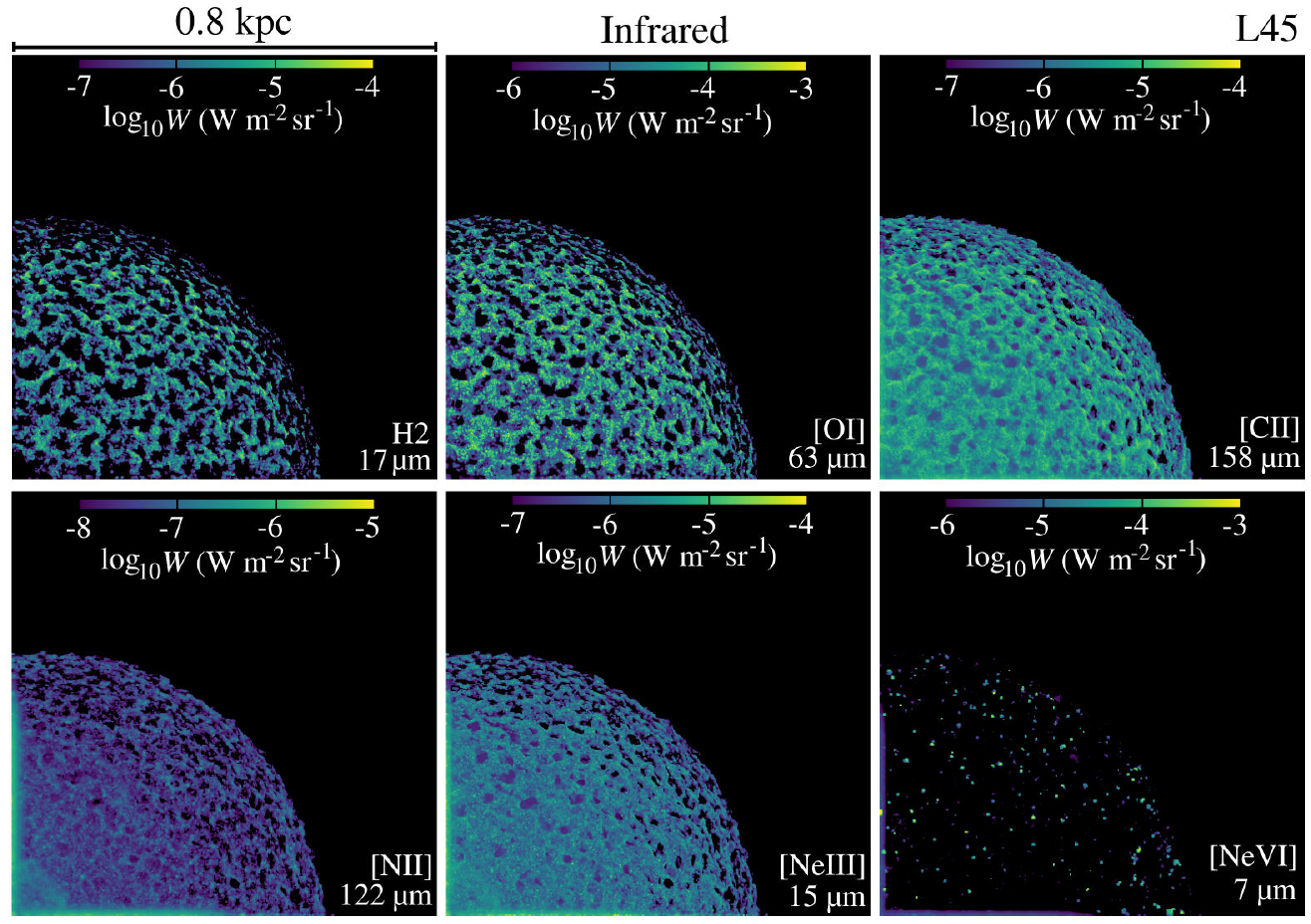}}
\mbox{
	\includegraphics[width=84mm]{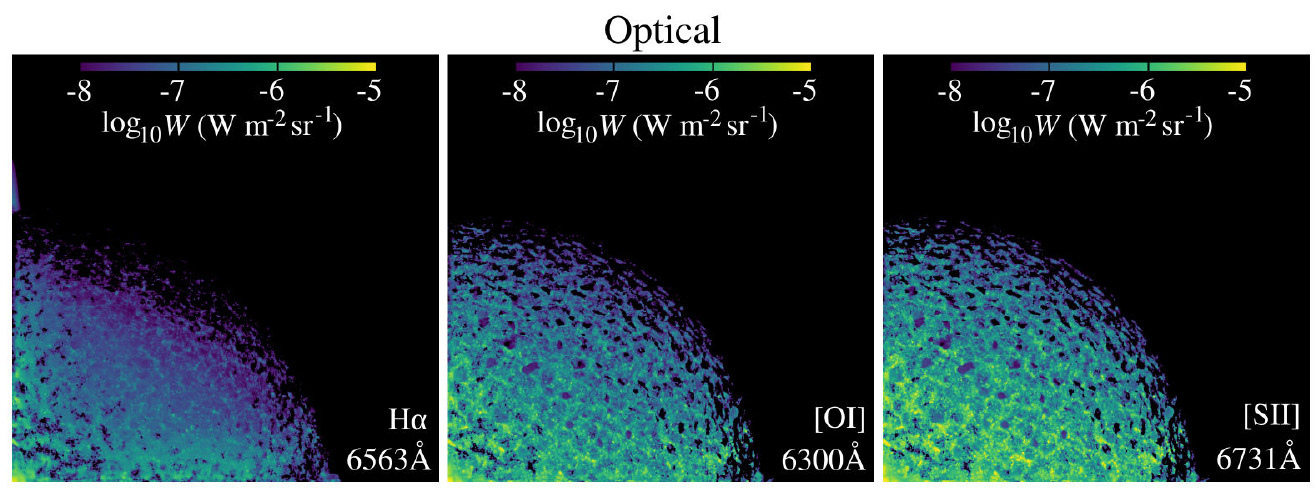}}
\vspace{-0.15 in}
\caption{Velocity-integrated maps of the continuum-subtracted infrared and optical emission lines from simulation L45, excluding the velocity ranges indicated by the grey bands in Fig.~\ref{spectra} to avoid emission from the ambient ISM. Only a subset of lines is shown here; the full sample can be found in Appendix~\ref{full_set_sect}. Each panel is 0.8~kpc across. The fluxes have been normalised to a distance of 184~Mpc, i.e. the luminosity distance to the quasar Mrk 231. In general, molecular and low-ionisation species trace clumpy structures in the outflowing shell, while intermediate ions trace a more diffuse phase, and the highest ionisation states arise from small regions of bright emission.  
\vspace{-0.15 in}} 
\label{images_L45}
\end{figure}

\begin{figure}
\centering
\mbox{
	\includegraphics[width=84mm]{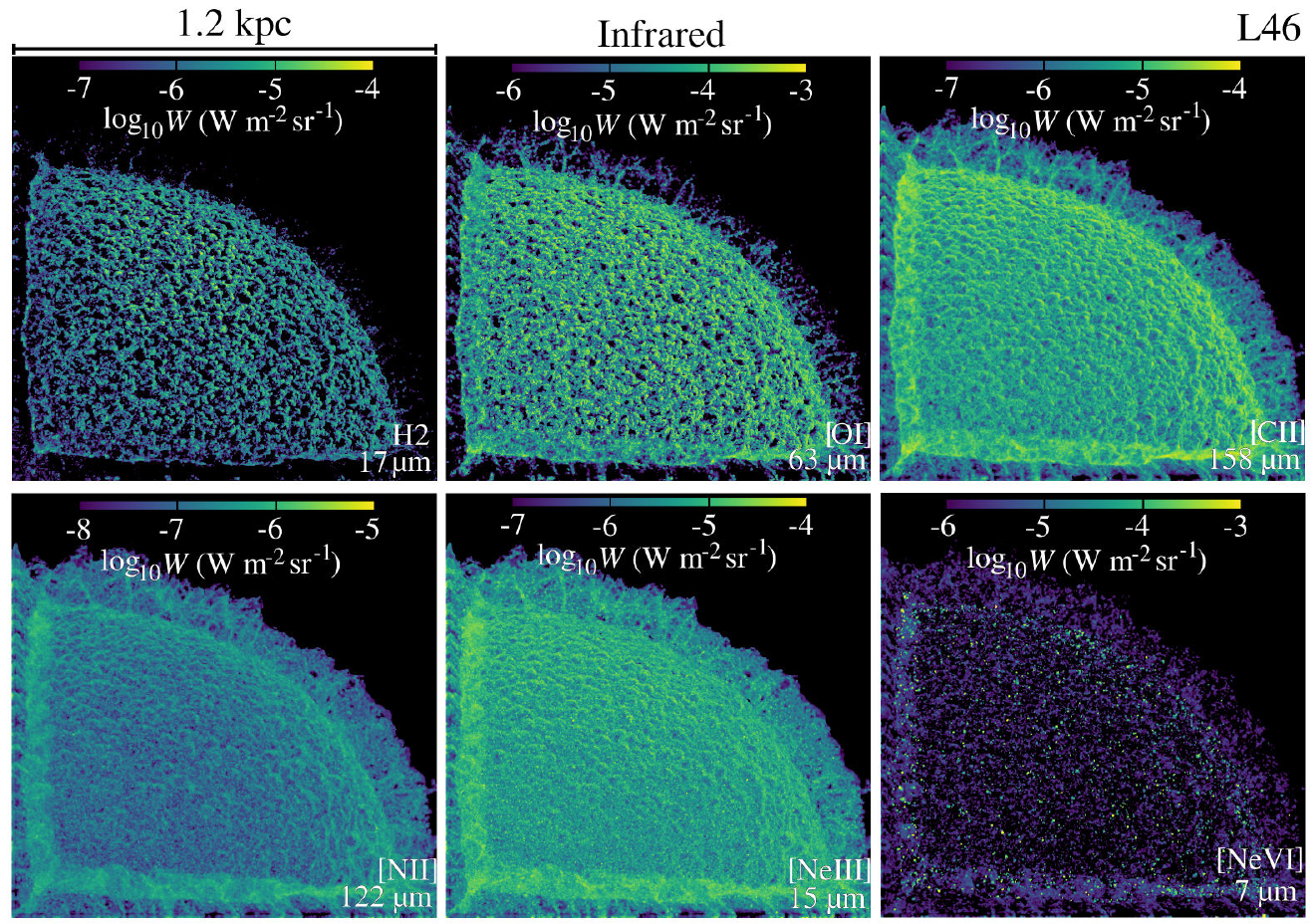}}
\mbox{
	\includegraphics[width=84mm]{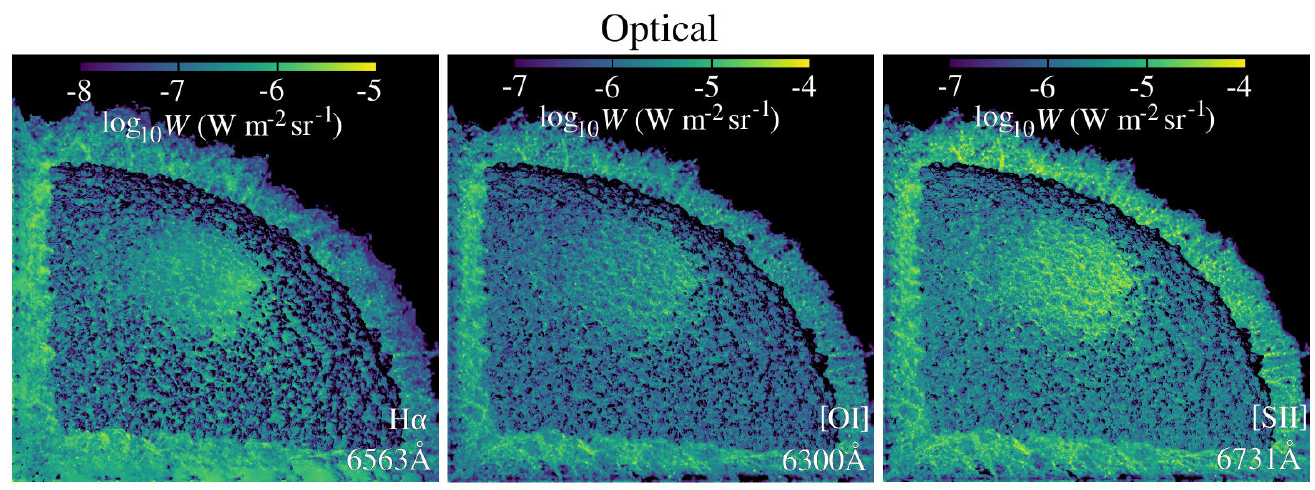}}
\vspace{-0.15 in}
\caption{As Fig.~\ref{images_L45}, but for simulation L46. Each panel is 1.2~kpc across. We again see the same trends in the spatial morphologies of the different species as for L45. 
\vspace{-0.15 in}} 
\label{images_L46}
\end{figure}

We can use these radiative transfer calculations to compare the spatial morphologies of each emission line from the outflow. In Figs.~\ref{images_L45} and \ref{images_L46} we show velocity-integrated maps of the infrared and optical emission lines, where we include emission from the red and blue wings (infrared) or the blue wing only (optical) as discussed above. The panels are each 0.8~kpc and 1.2~kpc across in L45 and L46, respectively. We again only show a subset of the lines studied in this paper; the full sample can be found in Appendix~\ref{full_set_sect}. 

The H$_{2, \, 17 \, \rm{\mu m}}$ line (top left panel of Figs.~\ref{images_L45} and \ref{images_L46}), which arises from the pure rotational S(1) transition of the H$_{2}$ molecule, traces dense, clumpy structures that have condensed from the cooling material within the outflowing shell. Comparing to the maps of CO emission (see Appendix~\ref{full_set_sect}), we find that the H$_{2, \, 17 \, \rm{\mu m}}$ line traces the same molecular structures as the CO emission. 

In the top middle panel, the infrared [O\textsc{i}]$_{63 \, \rm{\mu m}}$ line traces the same structures as the H$_{2, \, 17 \, \rm{\mu m}}$ emission. In contrast, the [O\textsc{i}]$_{6300 \, \text{\AA}}$ optical line is somewhat more diffuse, even though both are produced by neutral atomic oxygen. We will see in Section~\ref{properties_sect} that the optical emission is only present in the warmer gas phase ($\sim$10$^{4} \, \rm{K}$). This is unsurprising, as the 6300~\AA \, line requires a collisional excitation energy $E / k_{\rm{B}}$$\sim$10$^{4} \, \rm{K}$, while the $63 \, \rm{\mu m}$ line requires 100$\times$ less energy and can thus be excited in colder gas. 

The [C\textsc{ii}]$_{158 \, \rm{\mu m}}$ emission (top right) is strong in these clumpy structures, but there is also a significant contribution in the more diffuse regions between the dense clumps. The intermediate ions (e.g. [N\textsc{ii}]$_{122 \, \rm{\mu m}}$ and [Ne\textsc{iii}]$_{15 \, \rm{\mu m}}$; left-hand and centre panels of the middle row) predominantly arise from the diffuse, volume-filling phase in the outflowing shell, while the highest ionisation states such as [Ne\textsc{vi}]$_{7 \, \rm{\mu m}}$ (right-hand panel of the middle row) are dominated by small, bright knots of emission. We will show in the next section that these high ionisation states are produced in gas that is undergoing a period of rapid cooling. As such, any one particular patch of gas spends a very short time producing these high ionisation lines. The small, bright knots that we see in [Ne\textsc{vi}]$_{7 \, \rm{\mu m}}$ are therefore short flashes of emission as these compact regions transition through a rapid cooling phase. 

\vspBottom
  
\subsection{Physical properties of the line-emitting gas}\label{properties_sect}

\begin{figure}
\centering
\mbox{
  \includegraphics[width=84mm]{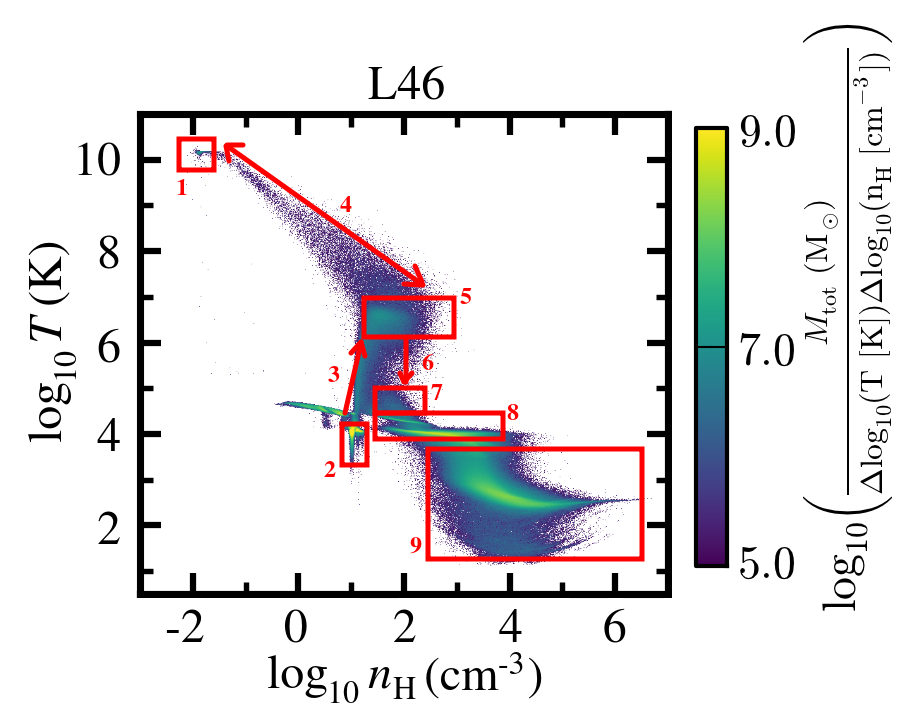}}
\vspace{-0.15 in}
\caption{Temperature versus density of all gas in the high-resolution octant of the L46 simulation after 1~Myr. The colour scale indicates the distribution of the total gas mass in this plane. We have highlighted several regions in this plot to guide the discussion. 
\vspace{-0.15 in}} 
\label{T_rho_tot}
\end{figure}

In this section we explore the physical properties of the gas probed by each emission line. Before we break down the gas properties by the separate emission line tracers, it is useful to look at the properties of the total gas distribution in the simulations. Fig.~\ref{T_rho_tot} shows the temperature versus density of all gas in the high-resolution octant of the L46 simulation, where the colour scale indicates the distribution of total gas mass in each pixel. We have highlighted several regions in this plot to guide the discussion. We only show L46 here, but the L45 simulation exhibits the same structures in temperature-density space (see also Fig.~5 of RFG18). The evolution of gas through the temperature-density plane proceeds as follows:

\begin{enumerate}[label=\arabic*),leftmargin=\parindent]
\item The ultra-fast outflow ($30 \, 000 \, \rm{km} \, \rm{s}^{-1}$) injected in the central parsec encounters the reverse shock (also called the wind shock) and is heated to $\sim$10$^{10} \, \rm{K}$, creating a hot bubble. The pressure of this hot bubble is $P_{\rm{hot}} / 2.3 k_{\rm{B}} = n_{\rm{H}} T = 2.0 \times 10^{8} \, \rm{cm}^{-3} \, \rm{K}$. This pressure accelerates the outflow, boosting the momentum beyond that expected in a momentum-driven outflow \citep{fauchergiguere12}. 
\item The ambient ISM is intially in thermal equilibrium at a density $n_{\rm{H}} = 10 \, \rm{cm}^{-3}$ before it has been hit by the outflow.
\item When the ambient ISM is swept up by the forward shock driven by the outflow, it is compressed and shock heated to $\sim$10$^{6.5 - 7} \, \rm{K}$. This corresponds to the post-shock temperature for a shock velocity of $\approx$$500 - 1000 \, \rm{km} \, \rm{s}^{-1}$, according to the Rankine-Hugoniot jump conditions for a strong shock, which is consistent with the outflow velocities that we find in the simulations. This gas is in pressure equilibrium with the hot bubble. 
\item Some of the material in the swept up shell mixes with the hot bubble, but remains in pressure equilibrium. 
\item The outflowing shell of gas swept up from the ambient medium initially remains close to the post-shock temperature.
\item Once the swept up gas reaches the peak in the cooling curve, the cooling time becomes very short and it rapidly cools. We can estimate the cooling time in this regime based on the analytic models presented in \citet{richings18b}. In Fig.~6 of that work, we presented the temperature evolution of the swept up shell of an isotropic AGN wind in a 1D analytic model based on \citet{fauchergiguere12}. The fiducial model (black curves) uses the same AGN luminosity, ambient ISM density and metallicity as our L46 simulation. In this model, the swept up shell cools from $10^{6} \, \rm{K}$ to $10^{4} \, \rm{K}$ in a time $t_{\rm{cool}} = 0.002 \, \rm{Myr}$. For comparison, the sound crossing time of the shell, with a thickness of $\approx$25~pc (from the simulations) and using the sound speed at $10^{6} \, \rm{K}$, is $t_{\rm{sound}} = 0.2 \, \rm{Myr}$. Hence the cooling time is 100$\times$ less than the sound crossing time in this regime, and so it cools isochorically, as the pressure of the surrounding hot medium has insufficient time to compress the cooling gas and maintain pressure equilibrium during this phase.
\item Gas that has just undergone rapid cooling and is approaching thermal equilibrium is now under-pressurised compared to the surrounding hot medium.
\item Once the gas reaches thermal equilibrium at a temperature $\sim$10$^{4} \, \rm{K}$, it is subsequently compressed, increasing in density and pressure. 
\item At higher densities, the gas undergoes a thermal instability and cools down to the cold phase at temperatures $\sim$100~K. The cooling in this regime proceeds isobarically, as the 10$^{4} \, \rm{K}$ and 100 K phases remain approximately in pressure equilibrium with one another, albeit with a large scatter. As we demonstrated in RFG18, this cold phase is conducive to the formation of new molecules. 
\end{enumerate}

\begin{figure}
\centering
\mbox{
  \includegraphics[width=82mm]{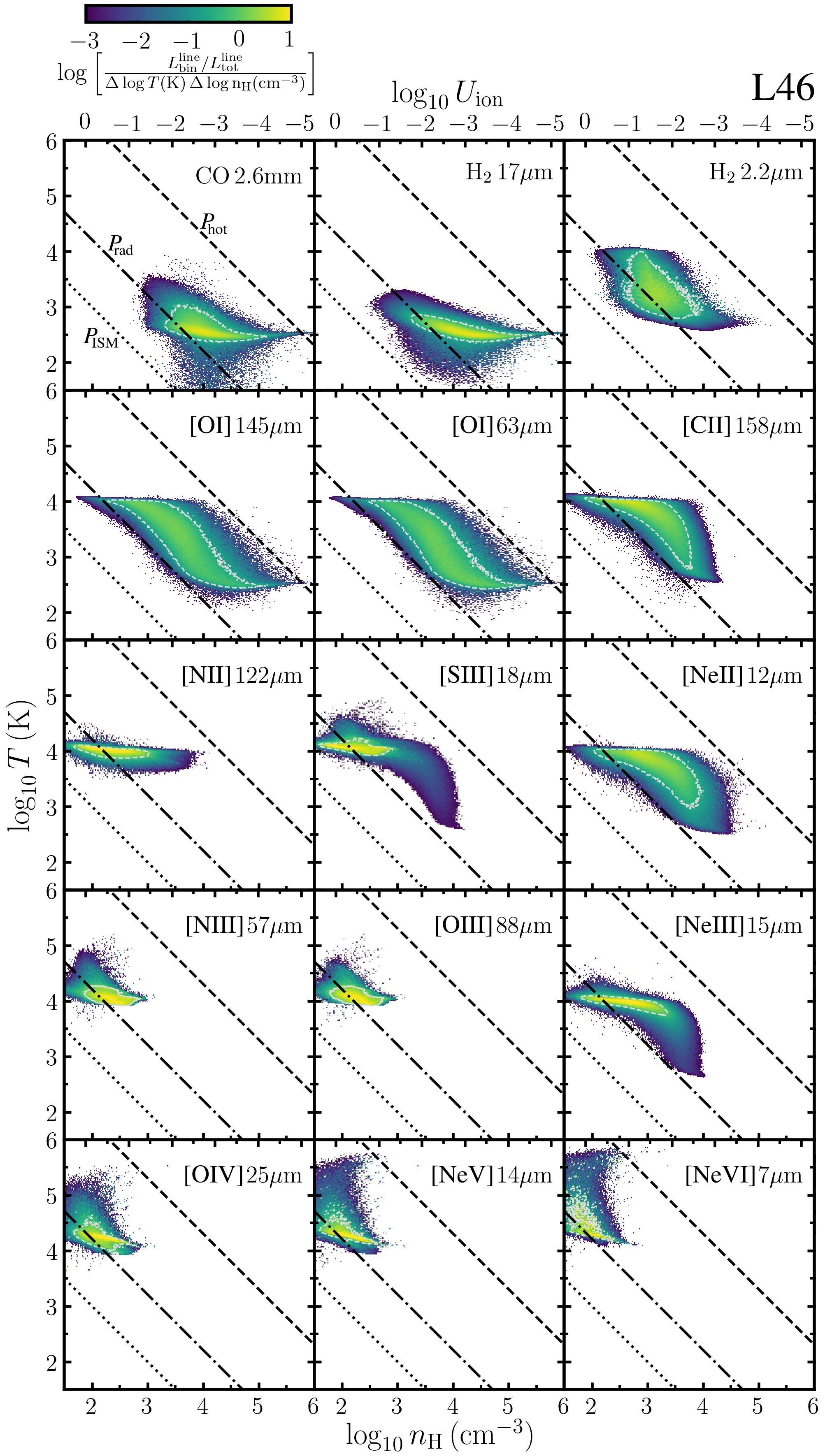}}
\vspace{-0.15 in}
\caption{Plots showing the temperature and density of gas producing the millimetre and infrared lines from the L46 simulation, arranged in order of increasing ionisation energy (or dissociation energy for molecules). The colour scale indicates the fraction of the total line luminosity that originates from gas in each 2D temperature-density bin. The dashed, dot-dashed and dotted lines show the pressure of the hot bubble ($P_{\rm{hot}}$, i.e. region (1) in Fig.~\ref{T_rho_tot}), the radiation pressure ($P_{\rm{rad}}$; see equation~\ref{P_rad_eqn}) at the radius of the outflowing shell (1.04~kpc), and the pressure of the ambient ISM ($P_{\rm{ISM}}$). The grey contours enclose 90 per cent of the total emission. The top axis shows the ionisation parameter, evaluated at 1.04~kpc. The hot bubble has compressed the line-emitting gas, resulting in higher pressures than in the ambient ISM or that would be produced from radiation pressure alone. However, this gas is under-pressurised compared to the hot bubble itself, especially for the highest ionisation states, which are located in gas that has recently undergone a rapid cooling phase. 
\vspace{-0.15 in}} 
\label{T_rho_IR_L46}
\end{figure}

\begin{figure}
\centering
\mbox{
  \includegraphics[width=82mm]{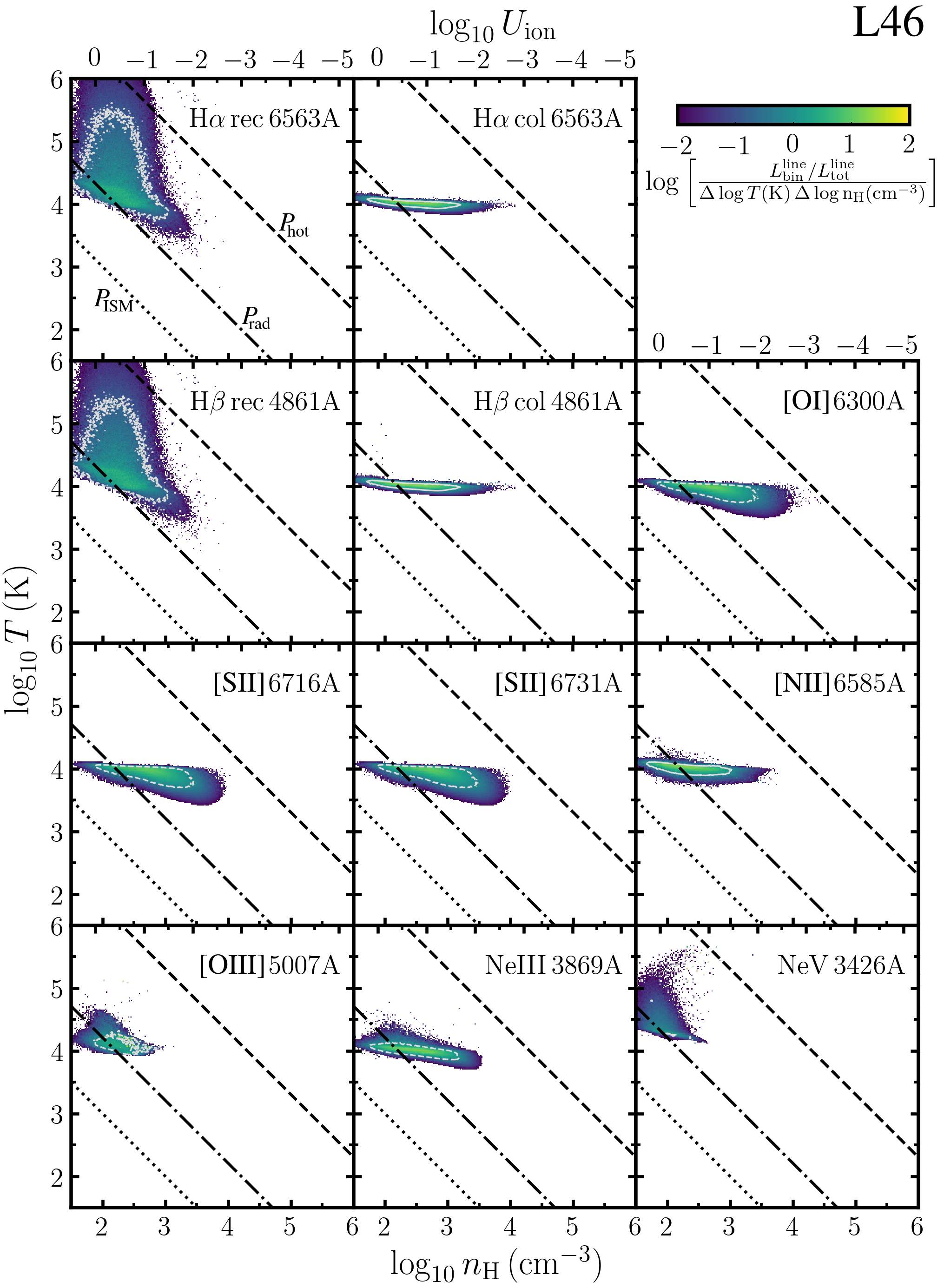}}
\vspace{-0.15 in}
\caption{As Fig.~\ref{T_rho_IR_L46}, but for the optical and UV emission lines. Like the millimetre and infrared lines, emission at optical and UV wavelengths also arises from gas that is over-pressurised compared to the ambient ISM and the radiation pressure alone, but under-pressurised compared to the hot phase. 
\vspace{-0.15 in}} 
\label{T_rho_optical_L46}
\end{figure}

Figs.~\ref{T_rho_IR_L46} and \ref{T_rho_optical_L46} show the temperature-density distribution of the line-emitting gas for the millimetre/infrared lines and the optical/UV lines, respectively. In these figures we show the full sample of 24 emission lines, not just a subset as in previous figures. The colour scale indicates the fraction of the total line luminosity that is produced from gas in each 2D temperature-density bin, normalised by the size of the bin in logarithmic temperature-density space. These are calculated using the individual line luminosities and ion/molecule-weighted temperatures and densities from each cell in the AMR grid used for the radiative transfer calculations (see Section~\ref{rt_sect}). We only show the L46 simulation here, but L45 exhibits the same trends. The grey contours in each panel enclose 90 per cent of the total line emission. 

In Fig.~\ref{T_rho_IR_L46}, we see in the top row that the CO$_{2.6 \, \rm{mm}}$ emission (from the $J$=1$-$0 transition) and the H$_{2, \, 17 \, \rm{\mu m}}$ emission (from the pure rotational S(1) transition) trace cold gas, at temperatures $\lesssim$10$^{3} \, \rm{K}$. In contrast, the  H$_{2, \, 2.2 \, \rm{\mu m}}$ line, from the v1$-$0 S(0) ro-vibrational transition, traces warmer molecular gas up to several thousand K. This is unsurprising given that the vibrational energy levels require a higher collisional excitation energy than the purely rotational transitions, and has also been noted many times elsewhere in the literature \citep[e.g.][]{veilleux20}. In observations of the Seyfert galaxy NGC 2110, \citet{rosario19} demonstrate that the warm molecular gas traced by the ro-vibrational v1$-$0 S(1) transition is spatially anti-correlated with the CO $J$=2$-$1 transition. This further supports the picture that the vibrationally excited H$_{2}$ and the CO trace different phases of molecular gas. 

The infrared lines of the low-ionisation states (e.g. O\textsc{i} and C\textsc{ii}) arise from gas that is cooling from the warm ($10^{4} \, \rm{K}$) to the cold ($100 \, \rm{K}$) phase, corresponding to region (9) in Fig.~\ref{T_rho_tot}. Intermediate ions such as N\textsc{ii} and O\textsc{iii} originate from gas that lies on the thermal equilibrium branch at 10$^{4}$~K, i.e. region (8) in Fig.~\ref{T_rho_tot}. Finally, the highest ionisation states such as Ne\textsc{v} and Ne\textsc{vi} are produced by gas that is coming to the end of the rapid cooling phase and is approaching the thermal equilibrium temperature.

In Fig.~\ref{T_rho_optical_L46} we have separated the hydrogen Balmer line emission between that produced by the recombination of H\textsc{ii} (`rec'; left panels of the top two rows) and that driven by the collisional excitation of H\textsc{i} (`col'; central panels of the top two rows). The recombination radiation arises from gas undergoing the rapid cooling phase, similarly to the high-ionisation states. In contrast, collisional excitation radiation arises only from neutral gas on the 10$^{4}$~K branch. \citet{raga15} demonstrated that, at 10$^{5}$~K, the emission coefficients of H$\alpha$ and H$\beta$ from collisional excitation are up to five orders of magnitude higher than those due to recombination. This can lead to significant contributions from collisional excitations at these high temperatures even for fairly low H\textsc{i} fractions. The lack of such collisional excitation radiation at 10$^{5}$~K in our simulations indicates that the H\textsc{i} fractions in the rapid cooling phase are very low, $<$10$^{-5}$. In L46, collisional excitation contributes 54 and 19 per cent to the total H$\alpha$ and H$\beta$ emission, respectively. In L45 (not shown), it contributes 43 and 10 per cent to H$\alpha$ and H$\beta$, respectively. 

The 6300~\AA \, emission from O\textsc{i} arises predominantly from the 10$^{4}$~K phase. This contrasts with the infrared lines from O\textsc{i}, which we saw in Fig.~\ref{T_rho_IR_L46} extends down to the cold phase. As we noted in Section~\ref{predict_sect}, this is as we would expect, as the 6300~\AA \, line requires a higher excitation energy. The optical and UV emission from S\textsc{ii}, N\textsc{ii}, O\textsc{iii} and Ne\textsc{iii} also trace the 10$^{4}$~K phase, while the Ne\textsc{v} UV emission includes gas towards the end of the rapid cooling phase.

To understand the role of photoionisation in the line-emitting gas, it is instructive to look at the ionisation parameter, $U_{\rm{ion}}$. This is defined as the ratio of the densities of hydrogen-ionising photons and hydrogen nuclei ($n_{\rm{\gamma}}$ and $n_{\rm{H}}$, respectively), and can be calculated as follows:

\begin{align}
  U_{\rm{ion}} &\equiv \frac{n_{\rm{\gamma}}}{n_{\rm{H}}} \\
               &= \frac{L_{\rm{ion}}^{13.6 \, \rm{eV} - 1 \, \rm{keV}}}{4 \pi r^{2} c n_{\rm{H}} \langle h \nu \rangle_{13.6 \, \rm{eV} - 1 \, \rm{keV}}}, 
\end{align}
where $L_{\rm{ion}}^{13.6 \, \rm{eV} - 1 \, \rm{keV}}$ is the ionising luminosity from 13.6~eV to 1~keV, $r$ is the distance from the AGN, $c$ is the speed of light, and $\langle h \nu \rangle_{13.6 \, \rm{eV} - 1 \, \rm{keV}}$ is the mean energy of ionising photons in the $13.6 \, \rm{eV} - 1 \, \rm{keV}$ band. We include ionising photons only up to 1~keV here, but the ionisation parameter would be almost unaffected if we instead included all photons $>$13.6~eV, as there are relatively few photons in the X-ray band. For the average quasar spectrum from \citet{sazonov04} that we use in our simulations, $\langle h \nu \rangle_{13.6 \, \rm{eV} - 1 \, \rm{keV}} = 32.3 \, \rm{eV}$, and $L_{\rm{ion}}^{13.6 \, \rm{eV} - 1 \, \rm{keV}} / L_{\rm{AGN}} = 0.10$, where $L_{\rm{AGN}}$ is the bolometric AGN luminosity. 

Since the line-emitting gas is located in a thin ($\approx$25~pc) spherical shell, it is all at approximately the same distance of $r = 1.04 \, \rm{kpc}$ (in L46) from the AGN. The ionising flux, and hence the density of ionising photons, is therefore approximately uniform for this gas, and thus $U_{\rm{ion}}$ depends simply on the inverse of the gas density. We therefore show $U_{\rm{ion}}$ at 1.04~kpc on the top axis of each panel in Figs.~\ref{T_rho_IR_L46} and \ref{T_rho_optical_L46}. In general, emission from the highest ionisation states arises from gas with $10^{-2}$$\lesssim$$U_{\rm{ion}}$$\lesssim$$10^{-1}$, while intermediate ions have $10^{-4}$$\lesssim$$U_{\rm{ion}}$$\lesssim$$10^{-2}$. Gas traced by molecules and low ionisation states extends to even lower ionisation parameters, down to $10^{-5}$. 

The dashed, dotted and dot-dashed lines in Figs.~\ref{T_rho_IR_L46} and \ref{T_rho_optical_L46} indicate the hot gas pressure ($P_{\rm{hot}}$, defined from region 1 in Fig.~\ref{T_rho_tot}), the initial pressure of the ambient ISM ($P_{\rm{ISM}}$) and the radiation pressure from the AGN at a radius of 1.04~kpc corresponding to the outflowing shell ($P_{\rm{rad}}$), respectively. The radiation pressure can be calculated from the ionising luminosity as follows:

\begin{equation}\label{P_rad_eqn} 
  P_{\rm{rad}} = \beta \frac{L_{\rm{ion}}^{^{13.6 \, \rm{eV} - 1 \, \rm{keV}}}}{4 \pi r^{2} c},
\end{equation}
where we again include ionising radiation only up to 1~keV, as the X-rays will penetrate beyond the ionised layer and therefore do not contribute to the radiation pressure here. The correction factor $\beta$, introduced by \citet{stern14a}, takes into account the contribution of non-ionising photons to the radiation pressure. They show that $\beta$$\approx$1 in dust-free gas, and $\beta$$\approx$2 in dusty gas. We showed in RFG18 that we require dust to be present in the outflowing shell to produce molecules consistent with observations. We therefore take $\beta = 2$ when calculating the radiation pressure. 

Note that, while photoionisation and photodissociation from the AGN radiation field are included in the thermo-chemistry in these simulations, direct radiation pressure itself is not included. This would tend to compress gas at thermal pressures less than $P_{\rm{rad}}$ \citep[e.g.][]{dopita02, draine11}. We see in Figs.~\ref{T_rho_IR_L46} and \ref{T_rho_optical_L46} that some of the emission lines arise from gas at pressures below $P_{\rm{rad}}$. We would therefore expect this gas to undergo additional compression if we were to include radiation pressure in the simulations, but the vast majority of emission in our simulations arises from gas with $P_{\rm{gas}}$$\gtrsim$$P_{\rm{rad}}$, and thus we do not expect the inclusion of radiation pressure in our simulations to significantly affect our results. 

We see that the hot bubble has compressed the line emitting gas to become over-pressurised compared to the ISM. For most emission lines, it also reaches higher pressures than would be achieved through direct radiation pressure alone. However, as this gas has cooled below the initial post-shock temperature of the outflowing shell, it is no longer in pressure equilibrium with the hot phase. The high ionisation species (e.g. Ne\textsc{vi}) trace gas that is most strongly under-pressurised compared to the hot medium, as it has not had time to be compressed and so is still at relatively low densities where higher ionisation states are prevalent. In contrast, the low- and intermediate-ionisation states trace gas at higher densities that have had longer to be compressed and are therefore closer to the pressure of the hot bubble, although they still remain under-pressurised. 

Photoionisation models of AGN emission lines such as \citet{groves04a} and \citet{stern14a} also assume thermal equilibrium in the line-emitting gas. Our simulations support this assumption for the intermediate lines such as [N\textsc{ii}]$_{122 \, \rm{\mu m}}$, [N\textsc{ii}]$_{6585 \, \text{\AA}}$, [S\textsc{ii}]$_{6716, \, 6731 \, \text{\AA}}$ and [O\textsc{i}]$_{6300 \, \text{\AA}}$. In Figs.~\ref{T_rho_IR_L46} and \ref{T_rho_optical_L46}, these lines trace a narrow region at the thermal equilibrium temperature $\sim$10$^{4} \, \rm{K}$. However, high ionisation states such as Ne\textsc{v} and Ne\textsc{vi} arise from a broad range of temperatures at a given density (spanning up to $\sim$0.5$-$1 dex) as they are coming to the end of the rapid cooling phase, and they have not yet reached thermal equilibrium. Emission from molecules and low-ionisation states such as [O\textsc{i}]$_{63, \, 145 \rm{\mu m}}$ and [C\textsc{ii}]$_{158 \, \rm{\mu m}}$ that trace the 100$-$10$^{4} \, \rm{K}$ phase also cover a broad range of temperatures at each density, rather than simply tracking the thermal equilibrium temperature as a function of density. This is likely due to turbulent shock heating. We find that, in our simulations, the dense ($n_{\rm{H}}$$>$$100 \, \rm{cm}^{-3}$) gas clouds that condense within the outflowing shell after it cools typically have velocity dispersions up to a few tens of km~s$^{-1}$. The shocks resulting from this turbulence can heat the gas up to a few thousand K. 

The photoionisation models also assume that the ions and molecules are in chemical equilibrium. To test this assumption in our simulations, we computed the equilibrium abundances of each gas particle at fixed density and temperature, and then repeated the radiative transfer calculations using the equilibrium abundances. The most notable difference is in the [O\textsc{iii}]$_{5007 \, \text{\AA}}$ line, which is $\approx$7$-$8 times higher when we set the abundances to equilibrium than when we use non-equilibrium abundances. Other emission lines also show modest deviations when we assume equilibrium abundances, in particular [Ne\textsc{v}]$_{14 \, \rm{\mu m}}$ (up to 80 per cent), [N\textsc{ii}]$_{122 \, \rm{\mu m}}$ (up to 62 per cent), [O\textsc{iv}]$_{25 \, \rm{\mu m}}$ (up to 60 per cent), [N\textsc{iii}]$_{57 \, \rm{\mu m}}$ (up to 44 per cent), [S\textsc{iii}]$_{18 \, \rm{\mu m}}$ (up to 39 per cent), [N\textsc{ii}]$_{6585 \, \text{\AA}}$ (up to 35 per cent), H$_{2; \, 2.2 \, \rm{\mu m}}$ (up to 31 per cent), H$_{2; \, 17 \, \rm{\mu m}}$ (up to 29 per cent), and CO$_{2.6 \, \rm{mm}}$ (up to 26 per cent). The remaining lines differ by $<$20 per cent when we use equilibrium abundances.

\vspBottom

\subsection{Comparison with observations}\label{comparison_sect}

\begin{figure}
\centering
\mbox{
	\includegraphics[width=84mm]{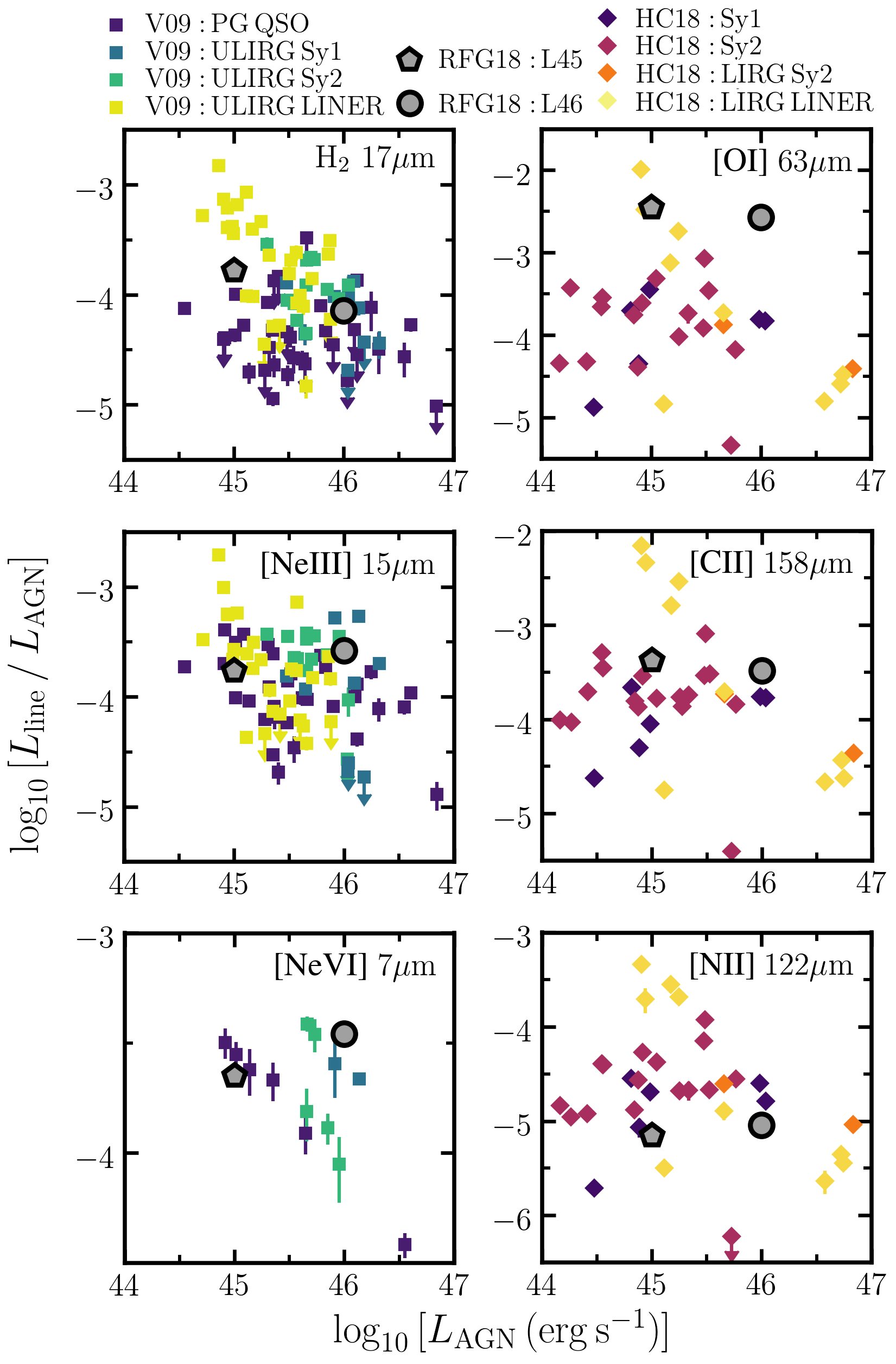}}
\vspace{-0.15 in}
\caption{Line luminosity ($L_{\rm{line}}$) divided by bolometric AGN luminosity ($L_{\rm{AGN}}$) plotted against $L_{\rm{AGN}}$. Our simulations from RFG18 are shown by the grey symbols. In the left column, showing a subset of mid-IR lines, we compare to the observations in \citet{veilleux09} (V09; squares), which we divide according to the AGN classification as PG QSO, ULIRG Sy1, ULIRG Sy2 and ULIRG LINER, as indicated by the colour. In the right column, showing a subset of far-IR lines, we compare to the observations in \citet{herreracamus18} (HC18; diamonds), divided according to classification as Sy1, Sy2, LIRG Sy2 and LIRG LINER (by colour). The full sample of all lines can be found in Appendix~\ref{full_set_sect}. While many of these lines from the simulations overlap with the observations, a notable exception is [O\textsc{i}]$_{63 \, \rm{\mu m}}$ (top right), for which the simulations overpredict the luminosity by an order of magnitude. 
\vspace{-0.15 in}} 
\label{IR_vs_Lagn}
\end{figure}

We now compare the predictions from our simulations to observations of outflows in AGN host galaxies. Fig.~\ref{IR_vs_Lagn} compares the  line luminosity ($L_{\rm{line}}$) divided by the bolometric AGN luminosity ($L_{\rm{AGN}}$) plotted against $L_{\rm{AGN}}$ for a subset of the infrared lines (the full sample can be found in Appendix~\ref{full_set_sect}). We divide by $L_{\rm{AGN}}$ on the y-axis to reduce the dynamic range; a simple linear scaling between $L_{\rm{line}}$ and $L_{\rm{AGN}}$ would thus manifest itself as a horizontal trend in these plots. For the simulations (RFG18; grey symbols), our radiative transfer calculations include only one quadrant of the spherical outflow (see Section~\ref{rt_sect}). We therefore multiply the resulting luminosities by four to obtain the luminosity of the whole outflow.

In the left-hand column of Fig.~\ref{IR_vs_Lagn} we compare mid-IR lines from our simulations to a sample of observed AGN host galaxies from \citet{veilleux09} (V09; squares). We divide the V09 sample according to the AGN classification as a Palomar-Green Quasar (PG QSO), UltraLuminous InfraRed Galaxy Seyfert 1 (ULIRG Sy1), ULIRG Sy2, or an ULIRG Low Ionisation Nuclear Emission Regions (LINER). We exclude galaxies classified as H\textsc{ii}-like. The different classifications are denoted by the colour. In the right-hand column we compare far-IR lines from our simulations to the observations from \citet{herreracamus18} (HC18; diamonds), divided as a Sy1, Sy2, LIRG Sy2 and LIRG LINER, again indicated by the colour. We exclude those classified as starburst galaxies in HC18. For the V09 sample, we use the bolometric luminosities listed in their work. However, this information is not included in HC18. For this latter sample we therefore use the bolometric luminosity given in V09 where available, otherwise we derive it from the far-IR (40~$\rm{\mu m}$$-$120~$\rm{\mu m}$) luminosity using the bolometric correction from Table~10 of V09 and assuming a 100 per cent AGN fraction, i.e. $L_{\rm{AGN}} = 1.22L_{\rm{FIR}}$.

For many of the lines shown in Fig.~\ref{IR_vs_Lagn}, the two simulations overlap with the observations. A notable exception is the [O\textsc{i}]$_{63 \, \rm{\mu m}}$ line. While there are two LIRG LINER systems that are close to, or higher than, the [O\textsc{i}]$_{63 \, \rm{\mu m}}$ luminosity predicted from L45, most of the observations lie at least an order of magnitude below the simulations.

Comparing to the individual AGN classifications, we see that in the H$_{2, \, 17 \, \rm{\mu m}}$ and [Ne\textsc{iii}]$_{15 \, \rm{\mu m}}$ lines the L45 simulation is closest to the ULIRG LINER systems, while L46 most closely overlaps with the ULIRG Sy1/Sy2 galaxies. In [Ne\textsc{vi}]$_{7 \, \rm{\mu m}}$ L46 is again closest to the ULIRG Sy1/Sy2, while L45 is closer to the PG QSOs. In the far-IR lines [C\textsc{ii}]$_{158 \, \rm{\mu m}}$ and [N\textsc{ii}]$_{122 \, \rm{\mu m}}$, it is more difficult to associate the simulations with a particular AGN classification. The observed LIRG LINERs bracket the line luminosities predicted by the simulations. However, in the observations these systems show a trend of decreasing $L_{\rm{line}} / L_{\rm{AGN}}$ with increasing $L_{\rm{AGN}}$, which is not replicated in the simulations. Most of the observed Sy1 and Sy2 galaxies have lower (higher) luminosities of [C\textsc{ii}]$_{158 \, \rm{\mu m}}$ ([N\textsc{ii}]$_{122 \, \rm{\mu m}}$) than the simulations, typically by a factor $\approx$3, although the most extreme examples of Sy1 and Sy2 galaxies in the observations bracket the simulations. 

\begin{figure}
\centering
\mbox{
	\includegraphics[width=84mm]{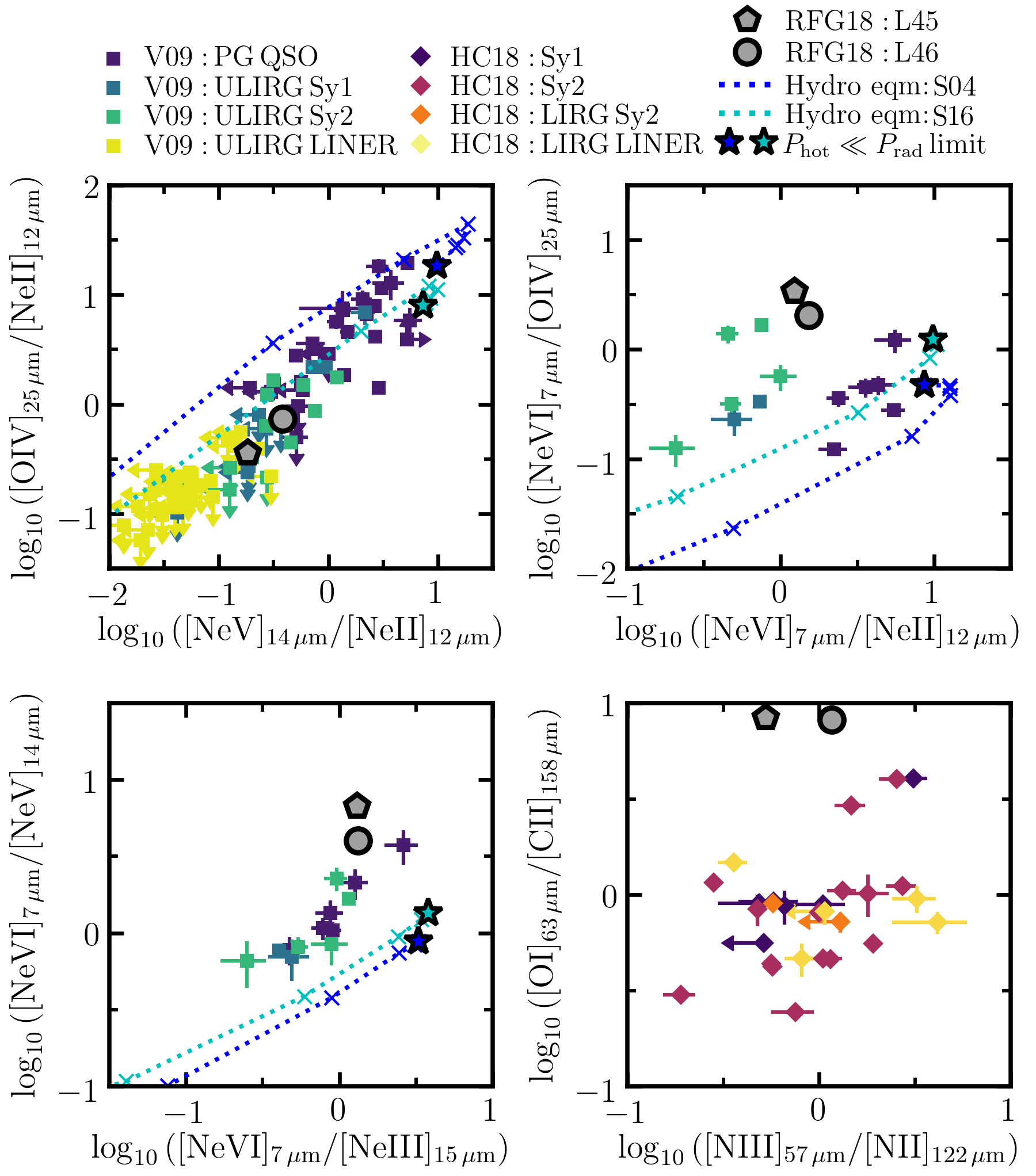}}
\vspace{-0.15 in}
\caption{Infrared line ratios from our simulations (RFG18; grey symbols) and observed AGN host galaxies from \citet{veilleux09} (V09; square symbols) and \citet{herreracamus18} (HC18; diamond symbols). The observed samples are again divided according to the AGN classification, as in Fig.~\ref{IR_vs_Lagn}. Upper and lower limits are denoted by arrows. The dotted lines show idealised \textsc{cloudy} photoionisation models in hydrostatic equilibrium based on \citet{stern16}, using the \citet{sazonov04} average quasar spectrum (S04; dark blue) and the fiducial \citet{stern16} model with an ionising slope of $-1.6$ (S16; cyan). The stars indicate the $P_{\rm{hot}}$$\ll$$P_{\rm{rad}}$ (i.e. RPC) limit of these models. In the top-left panel, the simulations are consistent with Sy2-like line ratios. However, the [Ne\textsc{vi}]$_{7 \, \rm{\mu m}}$ / [O\textsc{iv}]$_{25 \, \rm{\mu m}}$ and [Ne\textsc{vi}]$_{7 \, \rm{\mu m}}$ / [Ne\textsc{v}]$_{14 \, \rm{\mu m}}$ ratios are $\approx$3$\times$ higher in the simulations, while the [O\textsc{i}]$_{63 \, \rm{\mu m}}$ / [C\textsc{ii}]$_{158 \, \rm{\mu m}}$ ratios are $\approx$10$\times$ higher. 
\vspace{-0.15 in}} 
\label{IR_ratios}
\end{figure}

While many of the emission line luminosities predicted from the simulations overlap with the observations in Fig.~\ref{IR_vs_Lagn}, the large dynamic covered by the observations limits the constraining power in comparing the luminosities of individual lines. However, emission line ratios are often confined to a narrower range in observed systems, which provides tighter constraints on the models. In Fig.~\ref{IR_ratios} we therefore compare infrared line ratios from the simulations (grey symbols) and the observed samples from V09 (squares) and HC18 (diamonds). The dotted curves show idealised \textsc{cloudy} photoionisation models in hydrostatic equilibrium where we vary the ratio of $P_{\rm{hot}} / P_{\rm{rad}}$, based on the models of \citet{stern16}. The dark blue curves show a model using the same \citet{sazonov04} average quasar spectrum as used in our simulations, while the cyan curves show the fiducial \citet{stern16} model with an ionising slope of $-1.6$, typical of an unobscured AGN spectrum. The blue stars in Fig.~\ref{IR_ratios} indicate the $P_{\rm{hot}}$$\ll$$P_{\rm{rad}}$ limit of these models. As the \textsc{cloudy} calculations end at a temperature of 100~K, they do not capture the full [C\textsc{ii}] and [O\textsc{i}] emission in the cloud, so we do not include the \textsc{cloudy} model predictions in the bottom-left panel of Fig.~\ref{IR_ratios}. We discuss these models further in Section~\ref{drive_sect}. 

In the top-left panel of Fig.~\ref{IR_ratios}, the simulations lie on the observed correlation between these two line ratios. \citet{veilleux09} also demonstrated this correlation in the observations, which represents an excitation sequence extending from the ULIRG LINERs in the bottom left, through the ULIRG Sy2 and Sy1 galaxies and on to the PG QSOs in the top right. They note that this trend can be understood if [Ne\textsc{v}]$_{14 \, \rm{\mu m}}$ and [O\textsc{iv}]$_{25 \, \rm{\mu m}}$ are produced by the AGN while [Ne\textsc{ii}]$_{12 \, \rm{\mu m}}$ comes from starburst activity. For fixed [Ne\textsc{v}]$_{14 \, \rm{\mu m}}$ / [O\textsc{iv}]$_{25 \, \rm{\mu m}}$, varying the relative contribution of the starburst will move the ratios along the observed correlation, with the PG QSOs exhibiting the smallest contribution from the starburst, while the ULIRG LINERs contain the highest starburst contribution. Our simulations lie closest to the observed ULIRG Sy2 systems, although we do not include contributions from a starburst component, which would tend to move the simulations down and to the left in this plot.

In the two panels showing ratios including [Ne\textsc{vi}] (top-right and bottom-left), the simulations lie above the observations by $\approx$0.5 dex. This suggests that the [Ne\textsc{vi}]$_{7 \, \rm{\mu m}}$ line may be too strong in our simulations. The predicted [Ne\textsc{vi}]$_{7 \, \rm{\mu m}}$ / [Ne\textsc{ii}]$_{12 \, \rm{\mu m}}$ and [Ne\textsc{vi}]$_{7 \, \rm{\mu m}}$ / [Ne\textsc{iii}]$_{15 \, \rm{\mu m}}$ ratios from the simulations in these two panels overlap with the observations, however the observations cover almost two orders of magnitude and so would still be consistent with reducing the [Ne\textsc{vi}]$_{7 \, \rm{\mu m}}$ emission in the simulations by $\approx$0.5~dex. 

In the bottom right panel of Fig.~\ref{IR_ratios}, the simulations again deviate from the observations. While the [N\textsc{iii}]$_{57 \, \rm{\mu m}}$ / [N\textsc{ii}]$_{122 \, \rm{\mu m}}$ ratios from the simulations and observations overlap, [O\textsc{i}]$_{63 \, \rm{\mu m}}$ / [C\textsc{ii}]$_{158 \, \rm{\mu m}}$ is an order of magnitude too high in the simulations. As we saw in fig.~\ref{IR_vs_Lagn}, this is due to the [O\textsc{i}]$_{63 \, \rm{\mu m}}$ line, which is 10$\times$ too high in the simulations. 

\begin{figure}
\centering
\mbox{
  \includegraphics[width=84mm]{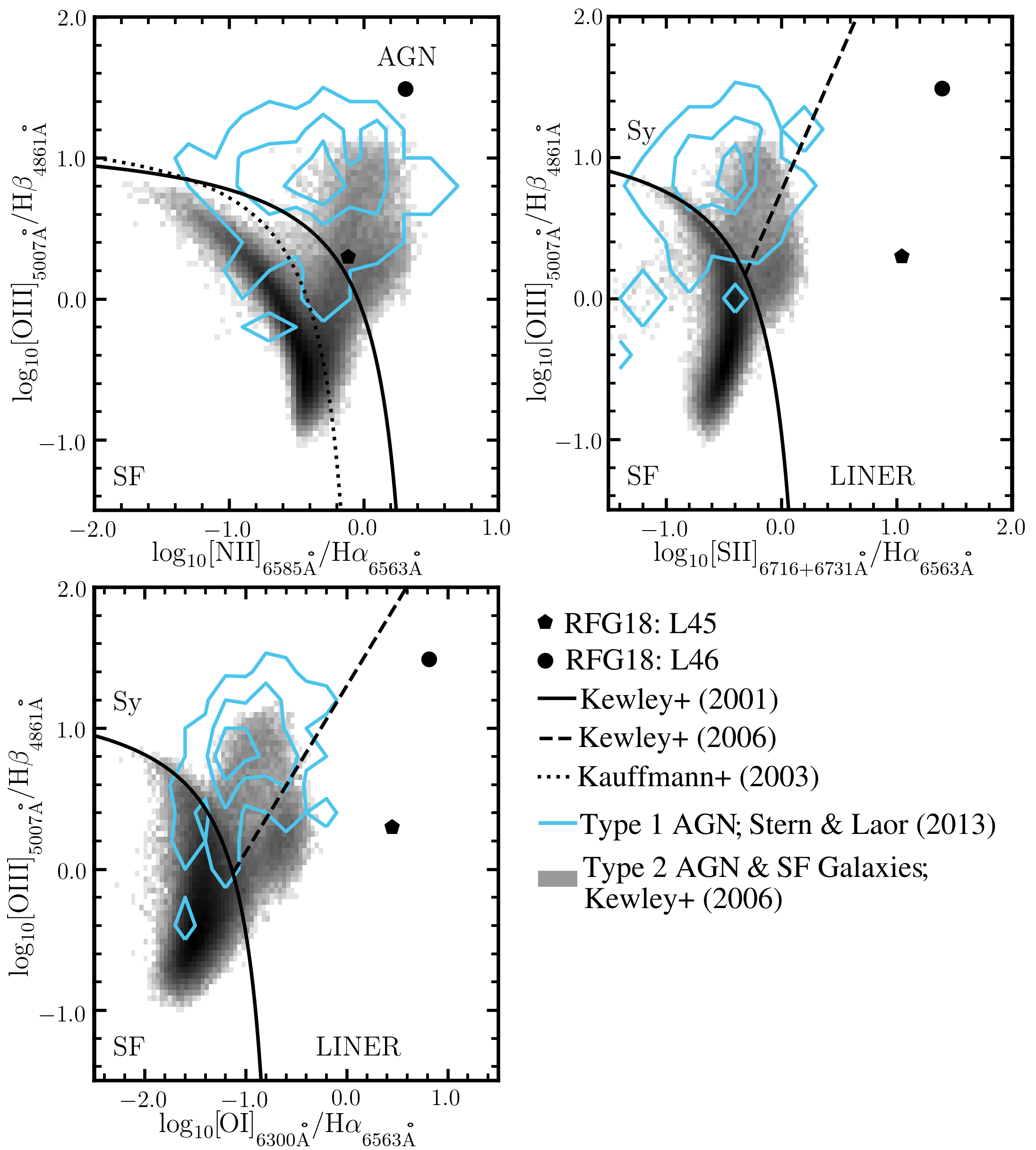}}
\vspace{-0.15 in}
\caption{BPT diagrams from our simulations (black symbols), star-forming galaxies and type 2 AGN from SDSS (\citealt{kewley06}; grey 2D histogram), and type 1 AGN from SDSS with bolometric luminosities $>$10$^{45} \, \rm{erg} \, \rm{s}^{-1}$ (\citealt{stern13}; blue contours). The black solid and dotted curves show the boundaries between star-forming and active galaxies from \citet{kewley01} and \citet{kauffmann03}, respectively, while the black dashed lines show the boundary between LINERs and Seyferts from \citet{kewley06}. The [S\textsc{ii}]$_{\rm{6716+6731 \, \text{\AA}}}$/H$\alpha$$_{\rm{6563 \, \text{\AA}}}$ and [O\textsc{i}]$_{\rm{6300 \, \text{\AA}}}$/H$\alpha$$_{\rm{6563 \, \text{\AA}}}$ line ratios are up to an order of magnitude higher in the simulations than is seen in the observations, due to strong X-ray heating in the assumed partially-obscured incident quasar spectrum and unresolved ionised layers. 
\vspace{-0.15 in}} 
\label{BPT_diagram}
\end{figure}

As noted in Section~\ref{rt_sect}, our radiative transfer calculations do not include dust attenuation from the ambient ISM of the host galaxy. While this has little effect on the infrared lines, the resulting luminosities of the optical and UV lines are much stronger than we would expect were such attenuation to be included. Nevertheless, we can compare the optical and UV line ratios to observations, as any additional dust attenuation would leave the ratios unaffected, provided that the wavelengths of the two lines are similar.

In Fig.~\ref{BPT_diagram} we show BPT diagrams \citep{baldwin81} of optical emission line ratios, which are commonly used to distinguish star-forming and AGN-dominated galaxies. The panels show the ratios of [N\textsc{ii}]$_{\rm{6585 \, \text{\AA}}}$ / H$\alpha$$_{\rm{6563 \, \text{\AA}}}$ (top left), [S\textsc{ii}]$_{\rm{6716+6731 \, \text{\AA}}}$ / H$\alpha$$_{\rm{6563 \, \text{\AA}}}$ (top right) and [O\textsc{i}]$_{\rm{6300 \, \text{\AA}}}$ / H$\alpha$$_{\rm{6563 \, \text{\AA}}}$ (bottom left) on the x-axis, plotted against [O\textsc{iii}]$_{\rm{5007 \, \text{\AA}}}$ / H$\beta$$_{\rm{4861 \, \text{\AA}}}$ on the y-axis in each case. The simulations are shown by the black symbols, while the grey 2D histograms show the distribution of star-forming galaxies and type 2 AGN in the Sloan Digital Sky Survey (SDSS) from \citet{kewley06} (updated to SDSS-DR8). The blue contours show observed type 1 AGN in SDSS from \citet{stern13}, for which we only include those with a bolometric luminosity $>$10$^{45} \, \rm{erg} \, \rm{s}^{-1}$ to select the most powerful AGN in this sample. The black solid and dotted curves show the boundaries between star-forming and active galaxies from \citet{kewley01} and \citet{kauffmann03}, respectively, while the black dashed lines show the boundary between LINERs and Seyferts from \citet{kewley06}.

In the top left panel panel of Fig.~\ref{BPT_diagram}, the L45 simulation overlaps with the observed AGN from SDSS and \citet{stern13}, although it lies close to the boundary between AGN and star-forming galaxies from \citet{kewley01}. L46 would also be classified as an AGN from this panel, although the [O\textsc{iii}]$_{\rm{5007 \, \text{\AA}}}$ / H$\beta$$_{\rm{4861 \, \text{\AA}}}$ is somewhat higher than is seen in even the most extreme AGN in the observations, by up to a factor $\approx$2.

In the top right and bottom left panels, the simulations lie on the LINER side of the classification boundaries. However, the [S\textsc{ii}]$_{\rm{6716+6731 \, \text{\AA}}}$ / H$\alpha$$_{\rm{6563 \, \text{\AA}}}$ and [O\textsc{i}]$_{\rm{6300 \, \text{\AA}}}$ / H$\alpha$$_{\rm{6563 \, \text{\AA}}}$ ratios are an order of magnitude higher than the observations. The H$\alpha$ emission traces recombinations driven by photoionisation, while the [S\textsc{ii}] and [O\textsc{i}] emission traces cooling in the Warm Neutral Medium (WNM) phase at $\sim$10$^{4} \, \rm{K}$ which is radiated via these metal lines. The high line ratios in the BPT diagrams predicted by the simulations therefore suggest that the ratio of the WNM to the photoionised phases is likely to be too high in our simulations, by up to an order of magnitude.

We find there are two effects that contribute to these anomalous ratios. Firstly, the WNM in our simulations is supported by X-ray heating. If we recompute the thermal equilibrium temperature without the X-ray component of the incident radiation field from the quasar, we find that much of the gas in the WNM cools from $10^{4} \, \rm{K}$ to a few thousand K. As noted in Section~\ref{sims_sect}, the average incident quasar spectrum that we use from \citet{sazonov04} has a lower UV to X-ray ratio, by a factor $\approx$2, than typical unobscured quasar spectra used in other AGN emission line models that successfully reproduce the BPT diagram \citep[e.g.][]{stern16}. This means there will be relatively stronger X-ray heating, compared to photoionisation driven by the UV band, which will increase the ratio of the WNM to photoionised phases and will thus contribute to the high line ratios that we find in the BPT diagrams.

Secondly, the transition from ionised to atomic hydrogen occurs at a column density $N_{\rm{H}, \, \rm{tot}}$$\approx$$10^{20.5} \, \rm{cm}^{-2}$ in our simulations. However, at a density of $n_{\rm{H}}$$=$$100 \, \rm{cm}^{-3}$ and our fiducial mass resolution of $m_{\rm{gas}}$$=$$30 \, \rm{M}_{\odot}$, this is comparable to the column density of a single gas particle, which suggests that the photoionised layer in our simulations is barely resolved. To estimate the extent to which we may be underestimating the photoionised component due to limited resolution, we can estimate the expected H$\alpha$ luminosity from recombinations of H\textsc{ii} given the incident ionising luminosity of the quasar. For the recombination emission coefficient that we use from \citet{raga15}, together with the rate coefficient for case B recombination used in \textsc{chimes}, we expect that each recombination of H\textsc{ii} will produce 0.27 H$\alpha$ photons at $10^{4} \, \rm{K}$. If we assume that every incident ionising photon is absorbed and thus leads to one recombination, the resulting H$\alpha$ luminosity that we would expect given the ionising luminosity is $\approx$3$-$4 times higher than that found in the simulations (before radiative transfer effects such as dust attenuation are included). This suggests that we underestimate the photoionised component by a factor $\approx$3$-$4.

We therefore conclude that these two effects of strong X-ray heating in the assumed partially obscured quasar spectrum and unresolved photoionised layers lead to the anomalously high line ratios in the BPT diagrams. Our simulations therefore do not provide reliable predictions for the BPT diagram. However, we caution that other emission line predictions presented above, in particular those comparing the photoionised and WNM phases, are also likely to be influenced by these effects. For example, the enhancement of the WNM due to strong X-ray heating will contribute to the anomalously high [O\textsc{i}]$_{63 \, \rm{\mu m}}$ luminosity seen in Fig.~\ref{IR_vs_Lagn}. 

\vspBottom

\section{Constraints on the driving mechanisms of AGN outflows}\label{drive_sect}

\begin{figure}
\centering
\mbox{
  \includegraphics[width=84mm]{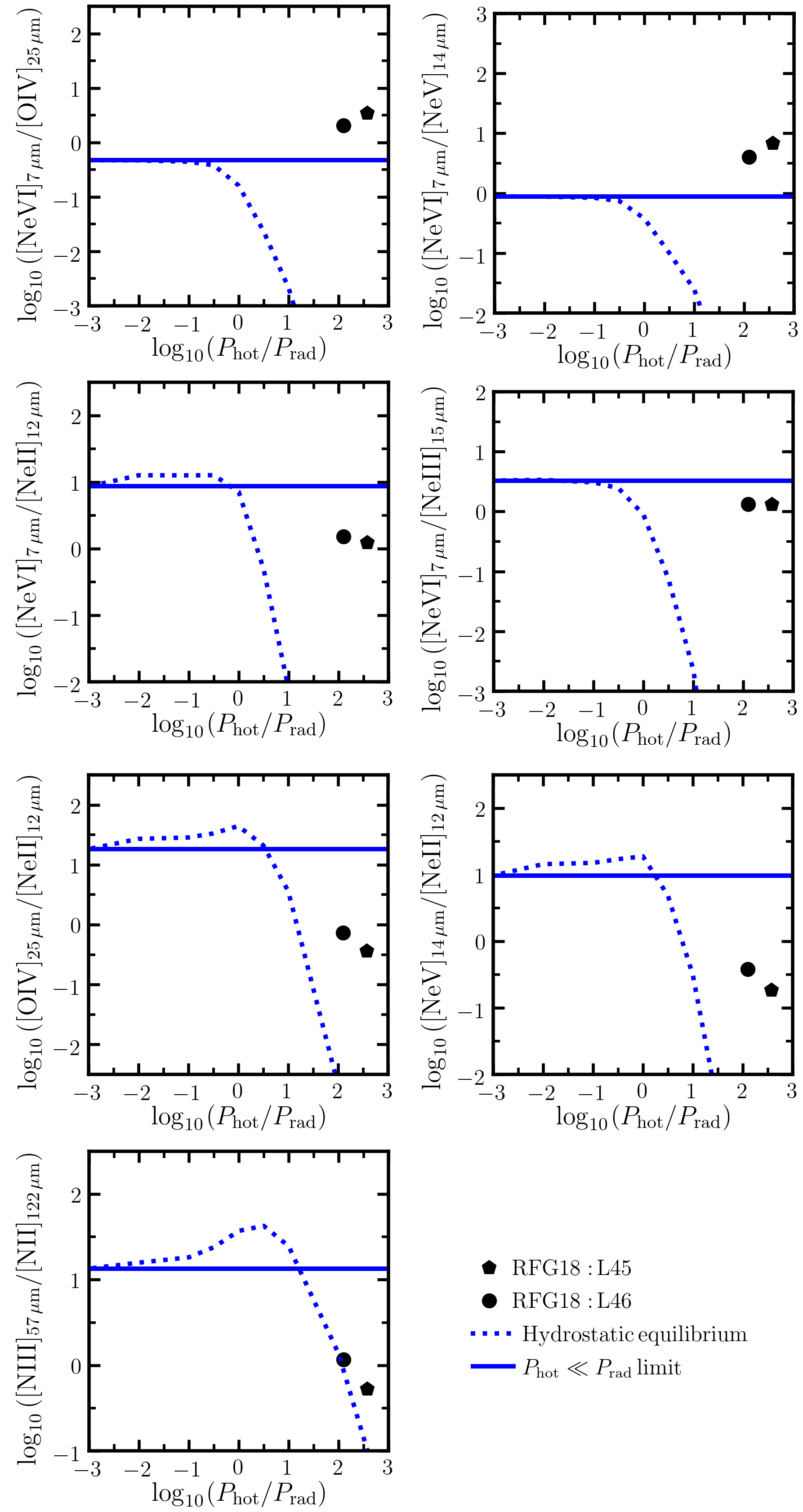}}
\vspace{-0.15 in}
\caption{Infrared line ratios versus the ratio of hot-gas to radiation pressure ($P_{\rm{hot}} / P_{\rm{rad}}$) in the simulations (black symbols) and an idealised \textsc{cloudy} photoionisation model in hydrostatic equilibrium based on \citet{stern16} (blue dotted curves). The horizontal blue lines indicate the $P_{\rm{hot}}$$\ll$$P_{\rm{rad}}$ (i.e. RPC) limit of the \textsc{cloudy} model. In the top \textsc{four} panels, the line ratios from the simulations are close to the RPC limit. However, in the bottom three panels the simulations lie more than 1 dex below the RPC limit predictions. These three line ratios in the bottom panels therefore present the best opportunity to observationally distinguish between the hot gas pressure scenario and the RPC limit. 
\vspace{-0.15 in}} 
\label{s16_IR_ratios}
\end{figure}

\begin{figure}
\centering
\mbox{
  \includegraphics[width=84mm]{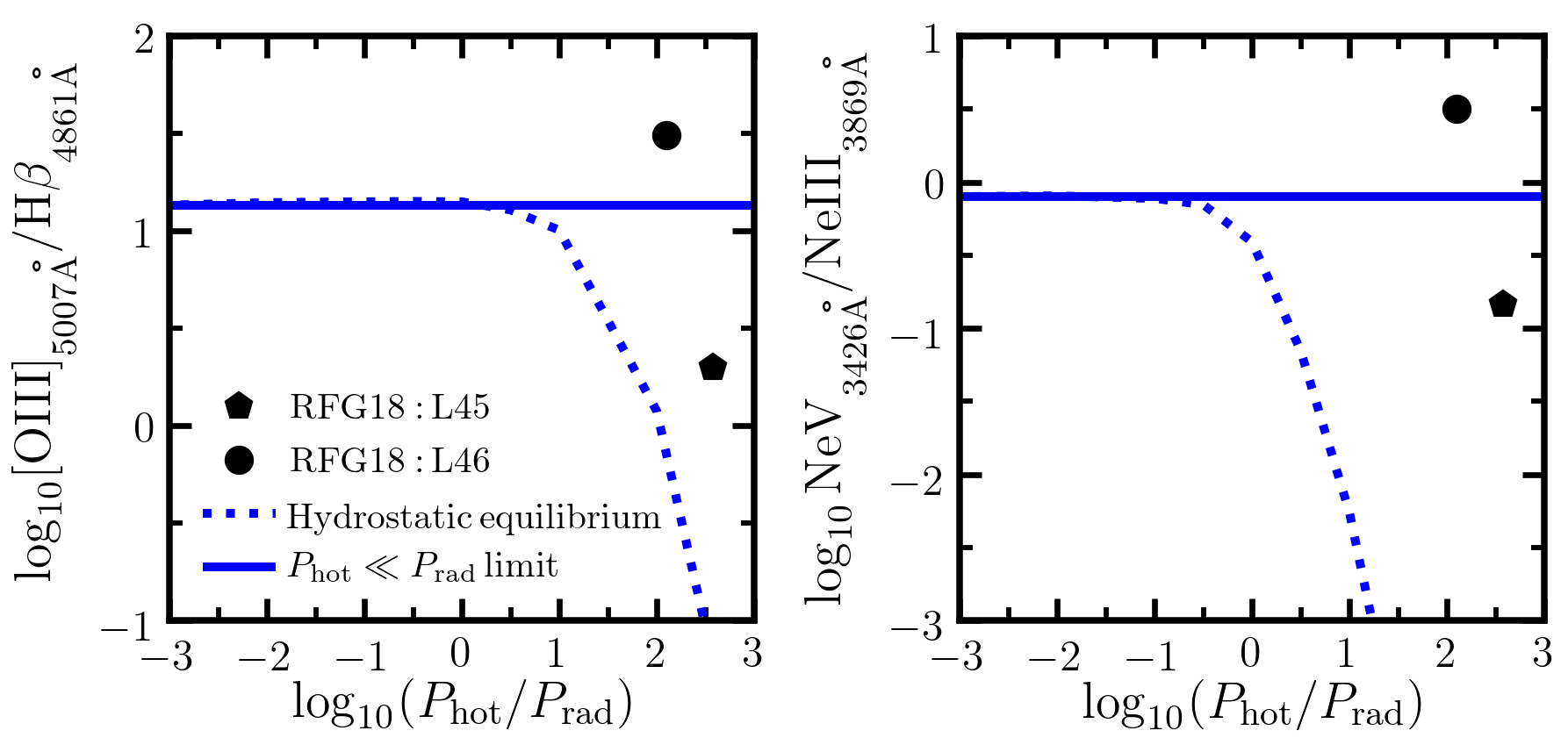}}
\vspace{-0.15 in}
\caption{The line ratios [O\textsc{iii}]$_{5007 \, \text{\AA}}$ / H$\beta$$_{\rm{4861 \, \text{\AA}}}$ (left) and Ne\textsc{v}$_{3426 \, \text{\AA}}$ / Ne\textsc{iii}$_{3869 \, \text{\AA}}$ (right) plotted against the ratio of hot gas pressure to radiation pressure ($P_{\rm{hot}} / P_{\rm{rad}}$). The black symbols show our simulations, while the blue dotted curves show the line ratios in an idealised \textsc{cloudy} photoionisation model in hydrostatic equilibrium based on \citet{stern16} using the \citet{sazonov04} average quasar spectrum. The line ratios in the $P_{\rm{hot}}$$\ll$$P_{\rm{rad}}$ (i.e. RPC) limit of the \textsc{cloudy} model are indicated by the horizontal blue lines. The line ratios from the simulations bracket the RPC limit of the \textsc{cloudy} model, which means that we cannot use these particular ratios to distinguish between the hot gas pressure dominated limit and RPC. 
\vspace{-0.15 in}} 
\label{s16_optical_ratios}
\end{figure}

\citet{stern16} argued that, in AGN outflows driven by the thermal pressure of a hot bubble, the line-emitting gas is compressed to higher pressures than we would expect in an outflow driven purely by the radiation pressure exerted by the AGN. They used idealised photoionisation calculations assuming hydrostatic balance between the hot phase and the line-emitting gas to show that this compression by the hot bubble leads to a dependence of the emission line ratios on the ratio of $P_{\rm{hot}}$ / $P_{\rm{rad}}$. By comparing these predictions to observations of quasar outflows, they concluded that, on large scales ($\gtrsim$1 kpc, comparable to the scales that we probe with our simulations), models with $P_{\rm{hot}} / P_{\rm{rad}} > 6$ are incompatible with the observed line ratios. This limits the dynamical role that hot gas pressure can play in driving galactic outflows in quasars. This comparison considered emission lines at optical and UV wavelengths, which selects unobscured quasars for which there are clear paths to the central AGN. As noted by \citet{stern16}, this selection might limit the effects of hot gas pressure as the hot bubble can begin to vent out and is no longer constrained by the dense ISM of the host galaxy \citep[e.g.][]{torrey20}. These results therefore do not rule out an earlier buried phase in which the hot gas pressure may dominate. 

While our simulations do include the photoionisation and photodissociation from the AGN radiation, we do not include the direct radiation pressure. However, we saw in Figs.~\ref{T_rho_IR_L46} and \ref{T_rho_optical_L46} that the radiation pressure is more than an order of magnitude less than the hot gas pressure in our simulations. We can therefore use our simulations to test the assumptions in the \citet{stern16} models in the $P_{\rm{hot}}$$\gg$$P_{\rm{rad}}$ regime. However, note that in our simulations the outflow remains constrained by the dense ISM, while \citet{stern16} focussed on comparisons with unobscured quasars in which the hot bubble may begin to vent out. 

As discussed in Section~\ref{properties_sect}, the hot bubble in our simulations does compress the line-emitting gas, as suggested in \citet{stern16}. However, the rapid cooling phase leaves this gas under-pressurised compared to $P_{\rm{hot}}$. We therefore cannot assume hydrostatic balance between the line-emitting gas and the hot phase. We explore the implications that this has for the emission line ratio constraints on $P_{\rm{hot}} / P_{\rm{rad}}$ below.

Fig.~\ref{s16_IR_ratios} shows infrared line ratios plotted against $P_{\rm{hot}} / P_{\rm{rad}}$, while Fig.~\ref{s16_optical_ratios} shows optical and UV line ratios versus $P_{\rm{hot}} / P_{\rm{rad}}$. The black symbols show our simulations, and the blue dotted curves show an idealised \textsc{cloudy} photoionisation model in hydrostatic equilibrium, including dust grains, based on the \citet{stern16} models but using the \citet{sazonov04} average quasar spectrum as in the hydrodynamic simulations. The horizontal blue lines indicate the Radiation Pressure Confinement (RPC) limit of the \textsc{cloudy} model, i.e. for $P_{\rm{hot}}$$\ll$$P_{\rm{rad}}$. 

The strongest distinction between our simulations and the RPC limit can be seen in the bottom three panels of Fig.~\ref{s16_IR_ratios}, for the infrared line ratios of [O\textsc{iv}]$_{25 \, \rm{\mu m}}$ / [Ne\textsc{ii}]$_{12 \, \rm{\mu m}}$, [Ne\textsc{v}]$_{14 \, \rm{\mu m}}$ / [Ne\textsc{ii}]$_{12 \, \rm{\mu m}}$ and [N\textsc{iii}]$_{57 \, \rm{\mu m}}$ / [N\textsc{ii}]$_{122 \, \rm{\mu m}}$. The predictions for these ratios from the simulation are more than 1~dex lower than the RPC limit of the \textsc{cloudy} model. We therefore conclude that these ratios present the strongest constraints between the hot gas pressure-driven and radiation pressure-driven scenarios. 

However, not all line ratios can be used to constrain the driving mechanism. In both panels of Fig.~\ref{s16_optical_ratios} the line ratio predictions from our simulations bracket the RPC limit of the idealised \textsc{cloudy} model, with L45 and L46 producing lower and higher ratios than in RPC, respectively. If we ran further simulations covering a wider range of the parameter space, we would expect some of these to produce intermediate ratios between the two simulations shown here, which could thus overlap with the RPC predictions. In the top four panels of Fig.~\ref{s16_IR_ratios}, the line ratios predicted by the simulations are also close to the RPC limit. These particular line ratios therefore cannot distinguish between the hot gas pressure-driven scenario and the RPC regime. This contrasts with the predictions of the \textsc{cloudy} model in the $P_{\rm{hot}}$$\gg$$P_{\rm{rad}}$ limit, in which these line ratios become much lower than in the RPC limit. This discrepancy between the simulations and the hot gas pressure-dominated limit of the \textsc{cloudy} model is due to the lower pressures that we find in the simulations compared to the assumption of hydrostatic balance with the hot phase. 

In general, the three infrared ratios most sensitive to the driving mechanism (i.e. the bottom three panels of Fig.~\ref{s16_IR_ratios}) compare high ionisation and low ionisation gas. We saw in Fig.~\ref{T_rho_IR_L46} that the high ionisation states are found in gas at lower densities that have had less time to undergo compression, while the low ionisation states trace gas that exhibit stronger compression and are thus at higher densities. Since the compression by the hot bubble produces the dense low-ionisation gas, the ratios between high and low ionisation states are most sensitive to the existence of the hot bubble and hence to $P_{\rm{hot}} / P_{\rm{rad}}$. However, line ratios such as [Ne\textsc{vi}]$_{7 \, \rm{\mu m}}$ / [Ne\textsc{iii}]$_{15 \, \rm{\mu m}}$ in the simulations are consistent with the RPC limit.

In Fig.~\ref{IR_ratios} we compared the infrared line ratios predicted by the \textsc{cloudy} photoionisation models in hydrostatic equilibrium to observations. The dark blue curves show the same models as in Fig.~\ref{s16_IR_ratios}, using the average quasar spectrum from \citet{sazonov04} as used in the simulations. We also show the fiducial model from \citet{stern16}, with an ionising slope of $-1.6$, which is typical for an unobscured quasar spectrum (cyan curves). The stars in Fig.~\ref{IR_ratios} show the RPC limit of the \textsc{cloudy} models. In the top two panels of Fig.~\ref{IR_ratios}, we see that the RPC limit predictions lie close to the Palomar-Green Quasars (PG QSO). However, in the bottom-left panel the RPC limit predictions are offset from the observations by $\approx$0.5 dex. As the \textsc{cloudy} calculations end at a temperature of 100~K, they do not fully capture the [C\textsc{ii}] and [O\textsc{ii}] emission, so we do not show these models in the bottom-right panel. Comparing the dark blue and cyan curves, we see that the assumed quasar spectrum can affect the line ratio predictions by up to a factor $\approx$2.

\vspBottom

\section{Energetics of different phases in the AGN outflow}\label{energetics_sect}

Emission lines are vital tools for measuring the energetics of galactic outflows. This is important for understanding how the AGN can influence its host galaxy, for example whether it can provide sufficient energy to unbind the gas in the galaxy and hence quench star formation \citep[e.g.][]{sturm11, cicone14, fluetsch19, lutz20}. However, we have seen in Section~\ref{properties_sect} that different lines trace different phases of the outflow. If we only measure the outflow energetics (e.g. mass outflow rate, momentum flux and energy flux) from a single line, we therefore would not capture the total energetics of the outflow.

\begin{figure*}
\centering
\mbox{
	\includegraphics[width=168mm]{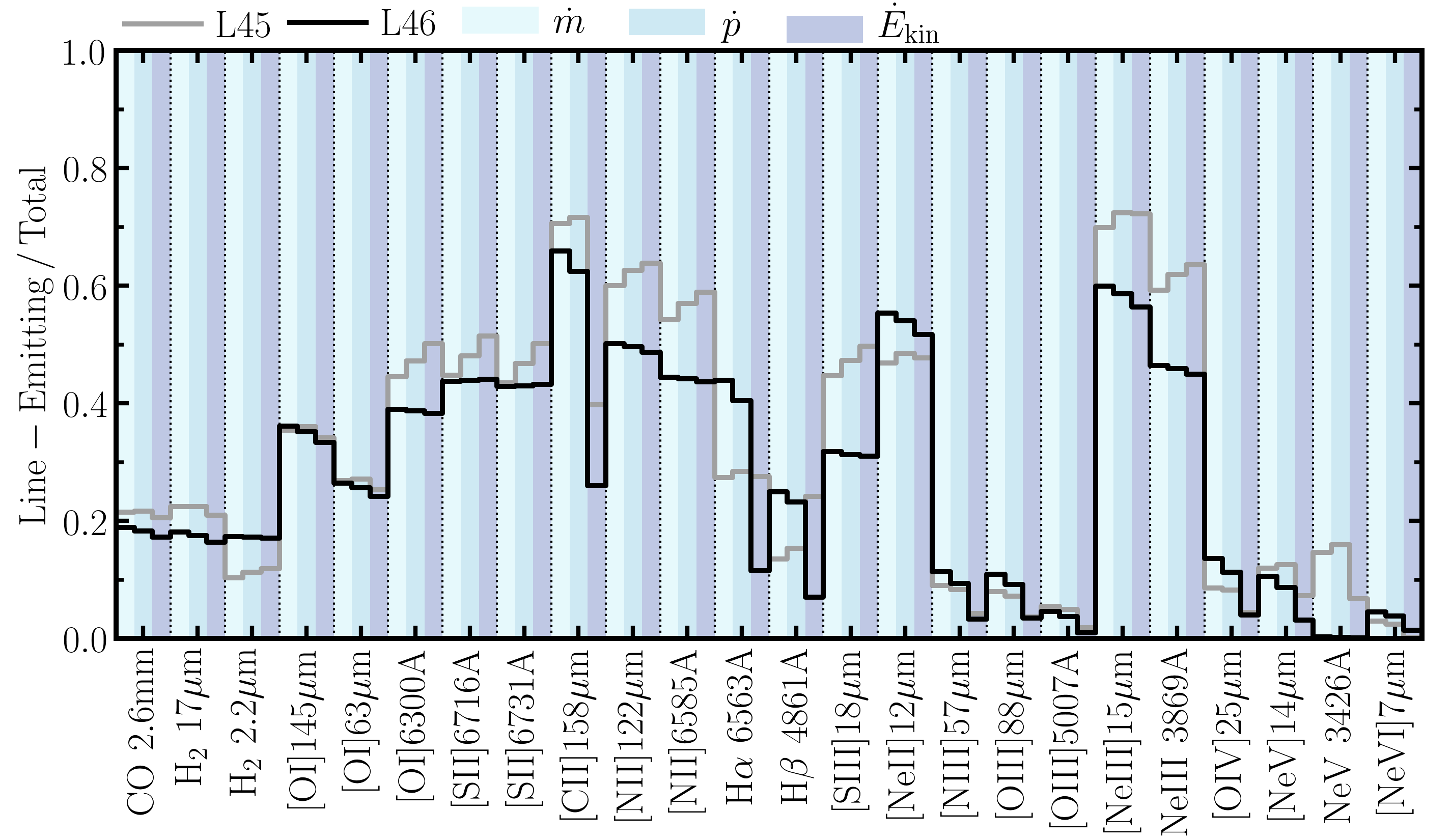}}
\vspace{-0.15 in}
\caption{Fractions of the total mass outflow rate ($\dot{m}$, light blue), momentum flux ($\dot{p}$, mid-blue) and kinetic energy flux ($\dot{E}_{\rm{kin}}$, dark blue) in gas traced by each emission line from the L45 (grey line) and L46 (black line) simulations, arranged in order of increasing ionisation (or dissociation) energy. The [C\textsc{ii}]$_{158 \, \rm{\mu m}}$, [N\textsc{ii}]$_{122 \, \rm{\mu m}}$ and [Ne\textsc{iii}]$_{15 \, \rm{\mu m}}$ lines each trace a significant fraction of the total energetics ($\gtrsim$50 per cent), while emission from high ionisation states such as [O\textsc{iv}]$_{25 \, \rm{\mu m}}$, [Ne\textsc{v}]$_{14 \, \rm{\mu m}}$ and [Ne\textsc{vi}]$_{7 \, \rm{\mu m}}$ traces much less ($\lesssim$10 per cent). 
\vspace{-0.15 in}} 
\label{energetics}
\end{figure*}

In this section, we investigate what fraction of the total energetics are captured by each emission line. We calculate this from the simulations as follows. First, we calculate the mass outflow rate ($\dot{m}_{\rm{part}}$), momentum flux ($\dot{p}_{\rm{part}}$) and kinetic energy flux ($\dot{E}_{\rm{kin, \, part}}$) of each gas particle in the high-resolution octant of the simulation volume. We define these quantities analogously to how they would be calculated in observations as follows:

\begin{equation}
  \dot{m}_{\rm{part}} = \frac{m_{\rm{part}} v_{\rm{los, \, part}}}{r_{\rm{part}}}.
\end{equation}

\begin{align}
  \dot{p}_{\rm{part}} &= \dot{m}_{\rm{part}} v_{\rm{los, \, part}} \nonumber \\
                      &= \frac{m_{\rm{part}} v_{\rm{los, \, part}}^{2}}{r_{\rm{part}}}.
\end{align}

\begin{align}
  \dot{E}_{\rm{kin, \, part}} &= \frac{1}{2} \dot{m}_{\rm{part}} v_{\rm{los, \, part}}^{2} \nonumber \\
                              &= \frac{1}{2} \frac{m_{\rm{part}} v_{\rm{los, \, part}}^{3}}{r_{\rm{part}}},
\end{align}
where $m_{\rm{part}}$, $v_{\rm{los, \, part}}$ and $r_{\rm{part}}$ are the particle mass, line-of-sight velocity and distance from the black hole, respectively. 

For each of the ion and molecule species, we then project these quantities on to the AMR grid used for the radiative transfer calculations weighted by the abundance of the given species. Next, we calculate the emissivity ($\epsilon$) of each emission line in every cell of the AMR grid (i.e. the line luminosity per unit volume). We then sort the cells in order of increasing $\epsilon$, and we find the emissivity $\epsilon_{90}$ for which 90 per cent of the total line luminosity orginates from cells with $\epsilon$$>$$\epsilon_{90}$. Finally, we calculate the $\dot{m}$, $\dot{p}$ and $\dot{E}_{\rm{kin}}$ summed over cells with $\epsilon$$>$$\epsilon_{90}$ for each line, which we compare to the same quantities summed over all cells to find the fraction of the total energetics that are traced by each line. 

Defined in this way, we are measuring the fraction of the energetics in cells that are `bright' in a given emission line (such that 90 per cent of the total emission is included). Note that a given cell can be bright in multiple lines, even from different ionisation states or molecules of the same element. For example, a cell containing a mixture of CO and C\textsc{ii} could be bright in both and thus would be accounted for in the energetics of each line; we do not divide a cell between the relative contributions to each line. Thus the fractions that we present here should not be considered as a division of the total energetics between each separate species. Rather, they tell us how much of the energetics can be captured by looking at each individual line, compared to how much arises from gas that is dark in the given line and thus would not be observable.

The results of this analysis are presented in Fig.~\ref{energetics}, which shows these fractions for each emission line. The species are arranged in order of increasing ionisation energy (ions) or dissociation energy (molecules) along the x-axis. For each species, we plot the fractions traced by the line-emitting gas for the mass outflow rate ($\dot{m}$), momentum flux ($\dot{p}$) and kinetic energy flux ($\dot{E}_{\rm{kin}}$), denoted by the background colour as light-, mid- and dark-blue, respectively. Results from the simulations L45 and L46 are shown by the grey and black lines, respectively.

We again caution that our predictions are likely to be influenced by uncertainties in the relative contributions of the photoionised and WNM phases, due to strong X-ray heating in the assumed partially obscured incident quasar spectrum and unresolved ionised layers, as discussed in Section~\ref{comparison_sect}. These results should therefore be considered as a qualitative indication of the relative importance of the different phases traced by each line, but the exact values remain uncertain. 

In general, the relative contributions of each line to the total momentum flux and kinetic energy flux are similar to the fraction of the total mass outflow rate that they trace. This is unsurprising given the idealised setup of these simulations, as most of the outflowing mass is located in a thin shell moving radially outwards with similar velocities across the shell. There are some exceptions particularly in $\dot{E}_{\rm{kin}} / \dot{E}_{\rm{kin, \, tot}}$, which can differ from $\dot{m} / \dot{m}_{\rm{tot}}$ and $\dot{p} / \dot{p}_{\rm{tot}}$ by factors of a few in some cases (e.g. [C\textsc{ii}]$_{158 \, \rm{\mu m}}$, [Ne\textsc{v}]$_{14 \, \rm{\mu m}}$ and [Ne\textsc{vi}]$_{7 \, \rm{\mu m}}$). The kinetic energy flux has the strongest dependence on velocity, so variations in the velocity distributions of different species will have a larger impact on $\dot{E}_{\rm{kin}} / \dot{E}_{\rm{kin, \, tot}}$ than the other quantities. Nevertheless, the trends that we see here are primarily driven by the fraction of the mass that is located in cells of the AMR grid that are bright in the given line.

The [C\textsc{ii}]$_{158 \, \rm{\mu m}}$ line (produced in the transition from $10^{4} \, \rm{K}$ to 100 K) and the [N\textsc{ii}]$_{122 \, \rm{\mu m}}$ and [Ne\textsc{iii}]$_{15 \, \rm{\mu m}}$ lines (produced in the $10^{4} \, \rm{K}$ phase) trace a large fraction of the mass outflow rate and momentum flux ($\approx$50$-$70 per cent). The latter two lines also capture most of the kinetic energy flux, although this is somewhat lower in [C\textsc{ii}]$_{158 \, \rm{\mu m}}$. It may seem surprising that [Ne\textsc{iii}] traces such a high fraction of the energetics, given that it has an ionisation energy close to species such as [O\textsc{iii}], which trace much lower fractions of the energetics. Comparing these emission lines in Figs.~\ref{T_rho_IR_L46} and \ref{T_rho_optical_L46}, we see this is because [Ne\textsc{iii}] emission extends to higher densities than the [O\textsc{iii}] lines. Photoionisation models of RPC clouds using \textsc{cloudy} also exhibit significant Ne\textsc{iii} abundances extending into the neutral region (see e.g. Fig.~A1 in \citealt{stern14a}). 

Emission from the high ionisation states, such as [O\textsc{iv}]$_{25 \, \rm{\mu m}}$, [Ne\textsc{v}]$_{14 \, \rm{\mu m}}$ and [Ne\textsc{vi}]$_{7 \, \rm{\mu m}}$, only trace low fractions of the total energetics (typically $\lesssim$10 per cent). As we saw in Section~\ref{properties_sect}, these species arise from gas in a transitionary phase as it reaches the end of a period of rapid cooling, before it is subsequently compressed to higher densities due to the external pressure exerted by the hot medium. Thus gas spends a relatively short period of time evolving through this phase.

We again stress that these results do not include dust attenuation from the host galaxy. This will not affect the millimetre and infrared lines, but we would expect those at optical and UV wavelengths to be strongly absorbed. The fractions presented in Fig.~\ref{energetics} for the optical and UV lines should therefore be interpreted as the fraction of the energetics in gas that emits these lines, but might not necessarily be observed if they are strongly attenuated by the host galaxy.

Observations of AGN- and star formation-driven outflows find prominent [C\textsc{ii}]$_{158 \, \rm{\mu m}}$ wings, both locally and at high redshift \citep[e.g.][]{maiolino12, cicone15, janssen16, bischetti19, herreracamus21}. This supports our prediction that the [C\textsc{ii}]$_{158 \, \rm{\mu m}}$ emission traces a large fraction of the outflow. \citet{fluetsch20} recently compared the relative contributions of the molecular, neutral atomic and ionised phases to the total mass outflow rate and energetics in local ULIRGs. The fractions that we present in Fig.~\ref{energetics} do not directly indicate the relative contributions of each species to the total. Instead, they show whether the emission from a given species is widely spread out over the whole outflow or concentrated in small regions. To compare our simulations to the data from \citet{fluetsch20}, we therefore also measured the relative contribution of the three phases from the fraction of the total outflowing mass in H$_{2}$, H\textsc{i} and H\textsc{ii}. We thus found that, in simulation L45 (L46), the relative mass fractions were as follows: molecular -- 25\% (17\%); neutral atomic -- 69\% (71\%); ionised -- 5\% (12\%). \citet{fluetsch20} also found a negligible contribution from the ionised phase, similar to our simulations. However, the outflows in \citet{fluetsch20} were dominated by the molecular phase, with on average more than 60\% of the mass outflow rate in H$_{2}$, and increasing even higher in AGN-dominated systems. In contrast, our simulations predict that the neutral atomic phase dominates. In RFG18 we found that the H$_{2}$ fraction in the simulations increases with increasing resolution, so it is possible that the simulations may approach the observed H$_{2}$ fractions at higher resolution. 

\vspBottom

\section{Conclusions}\label{conclusions_sect}

In this paper we explore the line emission from molecular, neutral atomic and ionised gas in simulations of multiphase, kiloparsec-scale outflows driven by the thermal pressure of a hot gas bubble generated by a central AGN. These simulations include an on-the-fly treatment for the non-equilibrium chemistry of ions and molecules coupled to the hydrodynamics and radiative cooling. The resulting ion and molecule abundances are used together with a radiative transfer code in post-processing to make predictions for the emission lines. We use these calculations to study how this emission traces the physical conditions and energetics of the outflow, and we compare the predicted emission lines from our models to observations. Our main results are as follows:

\begin{enumerate}[leftmargin=\parindent]
  \item We find that molecules (CO, H$_{2}$) and low-ionisation states (C\textsc{ii} and the infrared lines of O\textsc{i}) trace clumpy structures that have condensed within the outflowing shell as it cooled, while the intermediate species (e.g. N\textsc{ii}, S\textsc{iii}, Ne\textsc{iii}) arise from a more diffuse phase throughout the shell. Emission from the highest ionisation states (e.g. Ne\textsc{v}, Ne\textsc{vi}) is concentrated in small, bright knots, which are produced by regions that are passing through a period of rapid cooling. As such, any one particular region of gas spends a relatively short period of time in this phase, and thus this high ionisation emission appears as short, bright flashes concentrated in small regions (see Figs.~\ref{images_L45} and \ref{images_L46}).
  \item Comparing the temperature-density distribution traced by each emission line, we find that emission from high ionisation states (e.g. Ne\textsc{v}, Ne\textsc{vi}) is produced by gas that is coming to the end of a period of rapid cooling. Intermediate ions such as N\textsc{ii} and O\textsc{iii} arise from gas at the thermal equilibrium temperature of $\sim$10$^{4}$~K, while low-ionisation states (e.g. C\textsc{ii} and the infrared lines of O\textsc{i}) and molecules trace the transition from the warm ($\sim$10$^{4}$~K) to the cold ($\sim$100~K) phase (Figs.~\ref{T_rho_IR_L46} and \ref{T_rho_optical_L46}).
  \item The hot bubble compresses the line-emitting gas beyond the initial pressure of the ambient ISM by 1$-$2 orders of magnitude. For many emission lines, it also reaches higher pressures than would be achieved in an outflow driven and compressed by radiation pressure (Figs.~\ref{T_rho_IR_L46} and \ref{T_rho_optical_L46}). 
  \item The gas is under-pressurised compared to the pressure of the hot medium by more than an order of magnitude. Also, while the intermediate ions are in thermal equilibrium at $\sim$10$^{4}$~K, the high ionisation states are up to $\sim$1 dex higher than the thermal equilibrium temperature, as they are still cooling. The molecules and low-ionisation states trace a broad range of temperatures above the thermal equilibrium, due to turbulent shock heating. This has implications for photoionisation models that compute AGN emission line intensities assuming thermal and/or pressure equilibrium (Figs.~\ref{T_rho_IR_L46} and \ref{T_rho_optical_L46}). 
  \item While the luminosities of many of the infrared lines predicted from our simulations overlap with AGN observations, there are notable discrepancies, in particular in the [O\textsc{i}]$_{63 \, \rm{\mu m}}$ line, which is an order of magnitude too high in the simulations (Fig.~\ref{IR_vs_Lagn}). Some of the predicted line ratios at infrared wavelengths are also inconsistent with observations. For example, [Ne\textsc{vi}]$_{7 \, \rm{\mu m}}$ / [O\textsc{iv}]$_{25 \, \rm{\mu m}}$ and [Ne\textsc{vi}]$_{7 \, \rm{\mu m}}$ / [Ne\textsc{v}]$_{14 \, \rm{\mu m}}$ are 3$\times$ too high, while [O\textsc{i}]$_{63 \, \rm{\mu m}}$ / [C\textsc{ii}]$_{158 \, \rm{\mu m}}$ is 10$\times$ too high (Fig.~\ref{IR_ratios}). 
  \item At optical wavelengths, the BPT diagnostic diagrams show that the [S\textsc{ii}]$_{\rm{6716+6731 \, \text{\AA}}}$ / H$\alpha$$_{\rm{6563 \, \text{\AA}}}$ and [O\textsc{i}]$_{\rm{6300 \, \text{\AA}}}$ / H$\alpha$$_{\rm{6563 \, \text{\AA}}}$ ratios are $\approx$1 dex higher in the simulations than the observations (Fig.~\ref{BPT_diagram}). These anomalous ratios are due to two effects. Firstly, our assumed incident spectrum for a partially obscurred quasar has a $\approx$2$\times$ higher X-ray to UV ratio than a more typical unobscured spectrum, which enhances the Warm Neutral Medium (WNM) supported by X-ray heating. Secondly, we do not fully resolve photoionised layers, which results in an underestimate of the photoionised phase by a factor $\approx$3$-$4. Our simulations therefore do not provide a reliable prediction for the BPT diagrams. We also caution that these uncertainties may also affect other emission line predictions from our simulations, in particular those that compare the WNM and photoionised phases.
  \item As the line-emitting gas is compressed by the hot bubble, we find that certain line ratios are sensitive to the ratio of the hot gas pressure to radiation pressure, $P_{\rm{hot}} / P_{\rm{rad}}$, providing a constraint on the driving mechanism of AGN outflows, as suggested by the photoionisation models of \citet{stern16} that assumed thermal, chemical and hydrostatic equilibrium and an idealised slab geometry. We find that the ratios of [O\textsc{iv}]$_{25 \, \rm{\mu m}}$ / [Ne\textsc{ii}]$_{12 \, \rm{\mu m}}$, [Ne\textsc{v}]$_{14 \, \rm{\mu m}}$ / [Ne\textsc{ii}]$_{12 \, \rm{\mu m}}$ and [N\textsc{iii}]$_{122 \, \rm{\mu m}}$ / [N\textsc{ii}]$_{122 \, \rm{\mu m}}$ show the strongest distinction between our simulations (for which $P_{\rm{hot}}$$\gg$$P_{\rm{rad}}$) and the radiation pressure-dominated limit of \textsc{cloudy} photoionisation models in hydrostatic equilibrium (Fig.~\ref{s16_IR_ratios}). We therefore suggest that these line ratios provide the strongest constraints on the relative dynamical importance of radiation pressure and hot gas pressure on the outflow. 
  \item We quantify the fraction of the total mass outflow rate, momentum flux and kinetic energy flux of the outflow that is traced by each emission line. We find that the [N\textsc{ii}]$_{122 \, \rm{\mu m}}$ and [Ne\textsc{iii}]$_{15 \, \rm{\mu m}}$ lines (arising from the 10$^{4}$~K phase) trace $\approx$50$-$70 per cent of the totals in all three quantities. [C\textsc{ii}]$_{158 \, \rm{\mu m}}$ (arising from the transition from 10$^{4}$~K to 100~K) also traces  $\approx$60$-$70 per cent of the mass outflow rate and momentum flux, but only $\approx$30$-$40 per cent of the kinetic energy flux. Meanwhile, the high ionisation states such as [O\textsc{iv}]$_{25 \, \rm{\mu m}}$, [Ne\textsc{v}]$_{14 \, \rm{\mu m}}$ and [Ne\textsc{vi}]$_{7 \, \rm{\mu m}}$ (produced in a transitionary phase as the gas undergoes rapid cooling) trace $\lesssim$10 per cent of the energetics (Fig.~\ref{energetics}).
\end{enumerate} 

Our simulations demonstrate that, in outflows driven by the thermal pressure of a hot gas bubble (as can be produced by the shock heating of a fast accretion disk wind, for example identified observationally as a BAL or UFO), the line-emitting gas is compressed by the hot phase, which allows us to constrain the dynamical importance of the hot gas versus radiation pressure, provided that we consider line ratios that are most sensitive to this effect. We also find that much of the emission arises from gas in highly transitionary phases, in particular for high ionisation states such as Ne\textsc{v} and Ne\textsc{vi} that are produced towards the end of a rapid cooling phase, which leads to the line-emitting gas being out of thermal-, pressure- and chemical-equilibrium in many cases. This highlights the importance of simulations to capture these dynamical effects and their impact on emission line predictions.

Some of the emission line ratios predicted by our simulations are inconsistent with observations of all AGN subtypes, as noted above. However, in this study we only have two simulations, which cover a very limited range of the parameter space. These simulations also use an idealised setup, for example with an initially uniform ambient ISM that lacks turbulent gas structures or a realistic host galaxy geometry. The outflowing shell also remains constrained by the ambient ISM throughout the simulations, and so we cannot probe the regime after the outflow breaks out of the dense regions of the host galaxy. The tensions between our simulations and the observations are therefore insufficient evidence to rule out the hot gas pressure-driven scenario for kiloparsec-scale AGN outflows, as we cannot determine whether these tensions are a limitation of the model or if they are simply due to the limited range and idealised nature of the simulations.

To conclusively distinguish between possible driving mechanisms of AGN outflows, we would therefore need to run a more extensive suite of simulations covering a wide range of the parameter space and different physical conditions. The resulting emission line predictions can then be compared to observations to constrain the models, similar to how grids of photoionisation models have been employed to constrain such models in the past \citep[e.g.][]{groves04b, stern16}. With ongoing improvements to the computational codes used to run these simulations, we expect this to be achievable in the near future. 

\section*{Acknowledgements}

We thank the anonymous referee for their report, and we thank Roberto Maiolino for his comments on the manuscript. We are also grateful to Carlos Frenk and Richard Bower for useful discussions. AJR was supported by a COFUND/Durham Junior Research Fellowship under EU grant 609412; and by the Science and Technology Facilities Council [ST/P000541/1]. CAFG was supported by NSF through grants AST-1517491, AST-1715216, and CAREER award AST-1652522; by NASA through grant 17-ATP17-0067; and by a Cottrell Scholar Award and a Scialog Award from the Research Corporation for Science Advancement. JS was supported by a NASA grant through HST-GO-15935 and by the German Science Foundation via DIP grant STE 1869/2-1 GE 625/17-1 at Tel Aviv University. The simulations used in this work were run on the Stampede supercomputer at the Texas Advanced Computing Center (TACC) through allocations TG-AST160035 and TG-AST160059 granted by the Extreme Science and Engineering Discovery Environment (XSEDE), which is supported by NSF grant number ACI-154562; and the Pleiades supercomputer through allocation s1480, provided through the NASA Advanced Supercomputing (NAS) Division at Ames Research Center. This work used the DiRAC@Durham facility managed by the Institute for Computational Cosmology on behalf of the STFC DiRAC HPC Facility (www.dirac.ac.uk). The equipment was funded by BEIS capital funding via STFC capital grants ST/K00042X/1, ST/P002293/1, ST/R002371/1 and ST/S002502/1, Durham University and STFC operations grant ST/R000832/1. DiRAC is part of the National e-Infrastructure.

\section*{Data availability}

The data underlying this article will be shared on reasonable request to the corresponding author. A public version of the \textsc{gizmo} code is available at \url{http://www.tapir.caltech.edu/~phopkins/Site/GIZMO.html}. A public version of the \textsc{chimes} code is available at \url{https://richings.bitbucket.io/chimes/home.html}.

\vspBottom 

{}

\appendix 

\section{Full sample of emission line spectra}\label{full_set_sect}

In the main text of this manuscript we presented only a subset of the emission line spectra (Fig.~\ref{spectra}), velocity-integrated emission maps (Figs.~\ref{images_L45} and \ref{images_L46}) and comparisons of the infrared line luminosities with observations (Fig.~\ref{IR_vs_Lagn}), for the sake of brevity. In this appendix we present these figures for the full sample of all emission lines considered in this study.

\begin{figure}
\centering
\mbox{
	\includegraphics[width=84mm]{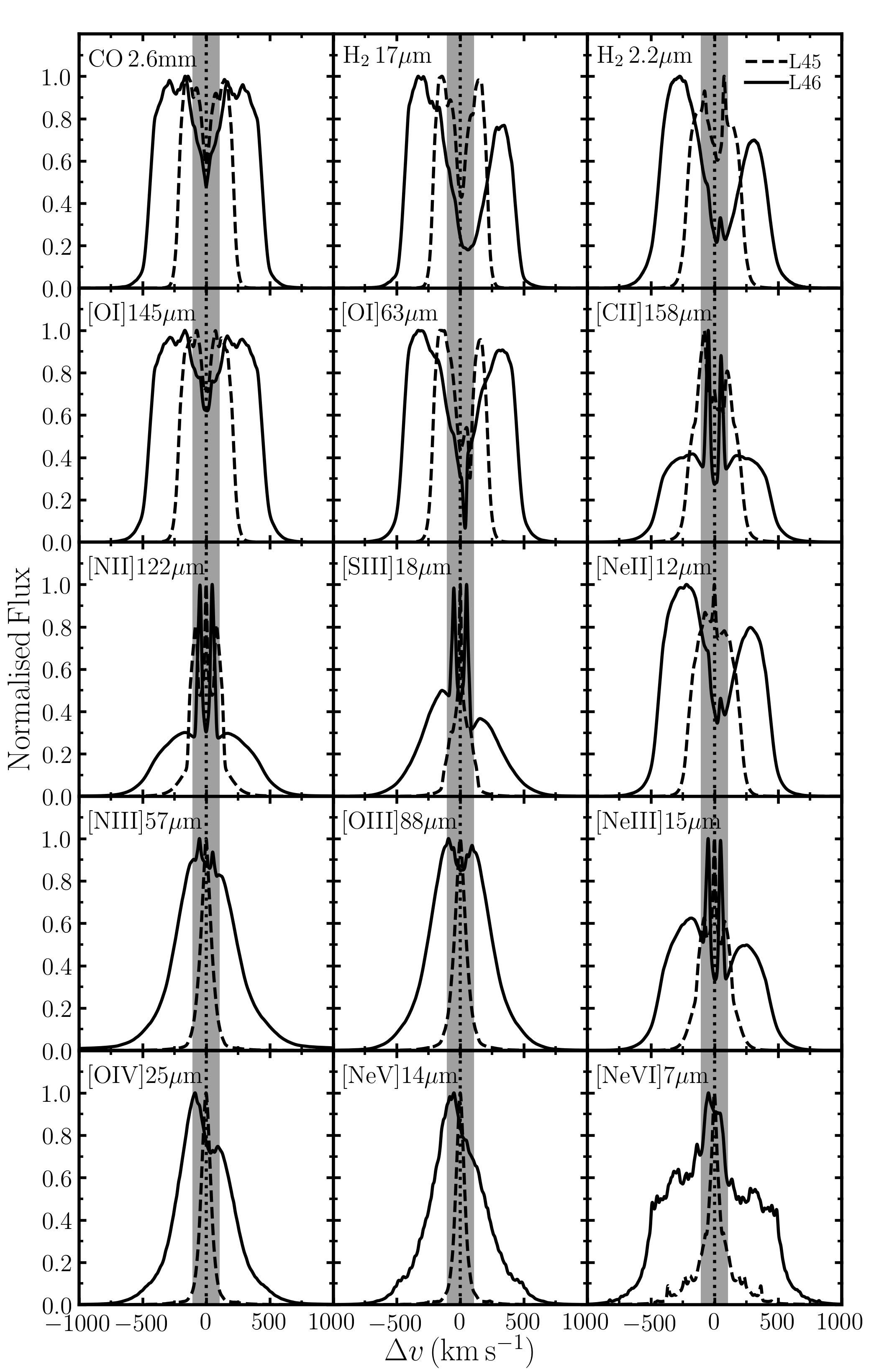}}
\mbox{
	\includegraphics[width=84mm]{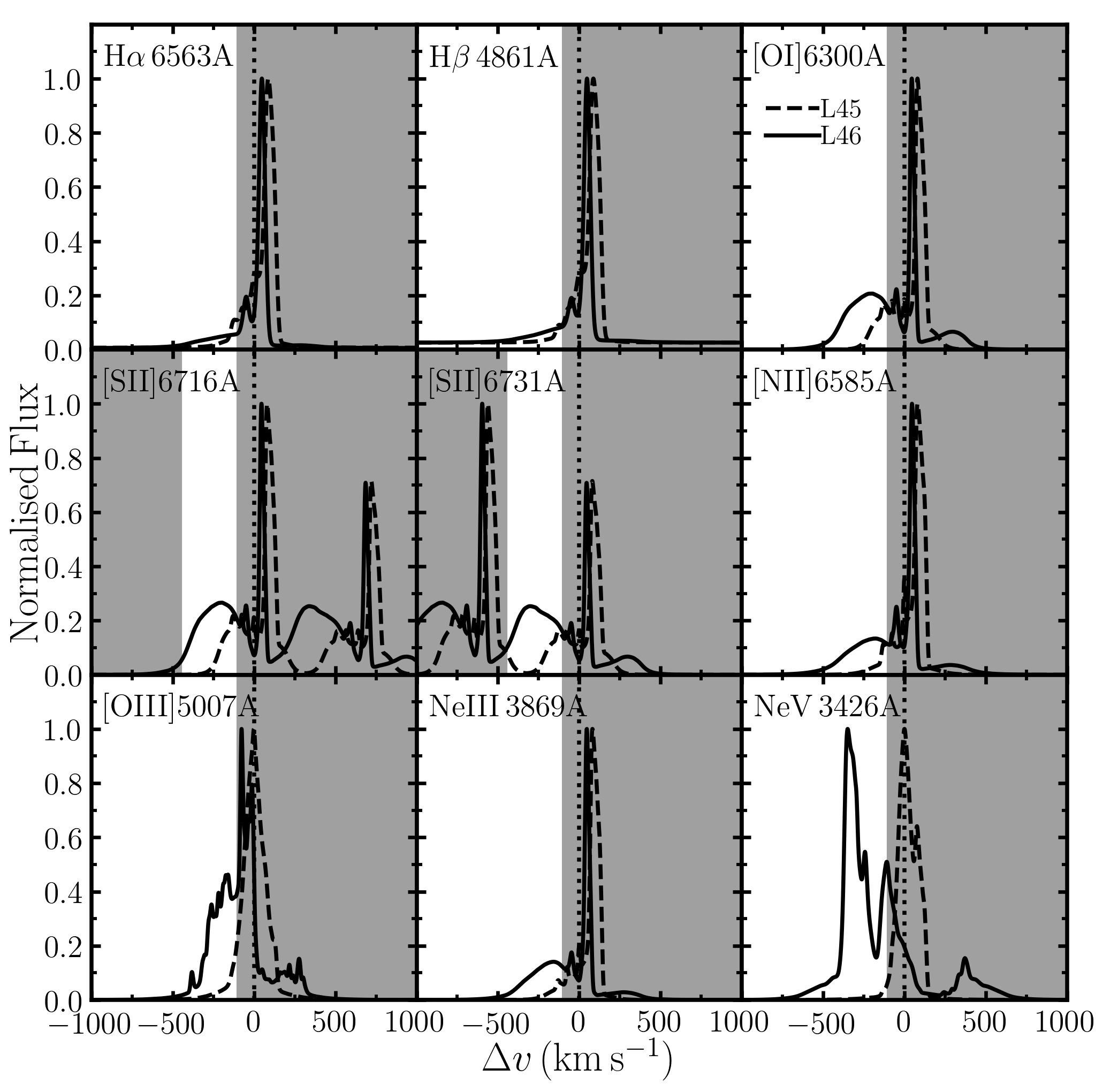}}
\vspace{-0.15 in}
\caption{Continuum-subtracted spectra of the millimetre and infrared emission lines (top panels) and the optical and UV lines (bottom panels) in the full sample, from simulations L45 (dashed curves) and L46 (solid curves). The grey shaded bands indicate velocity ranges that are excluded from further analysis, to focus on the outflow component.  
\vspace{-0.15 in}} 
\label{full_spectra}
\end{figure}

\begin{figure}
\centering
\mbox{
	\includegraphics[width=84mm]{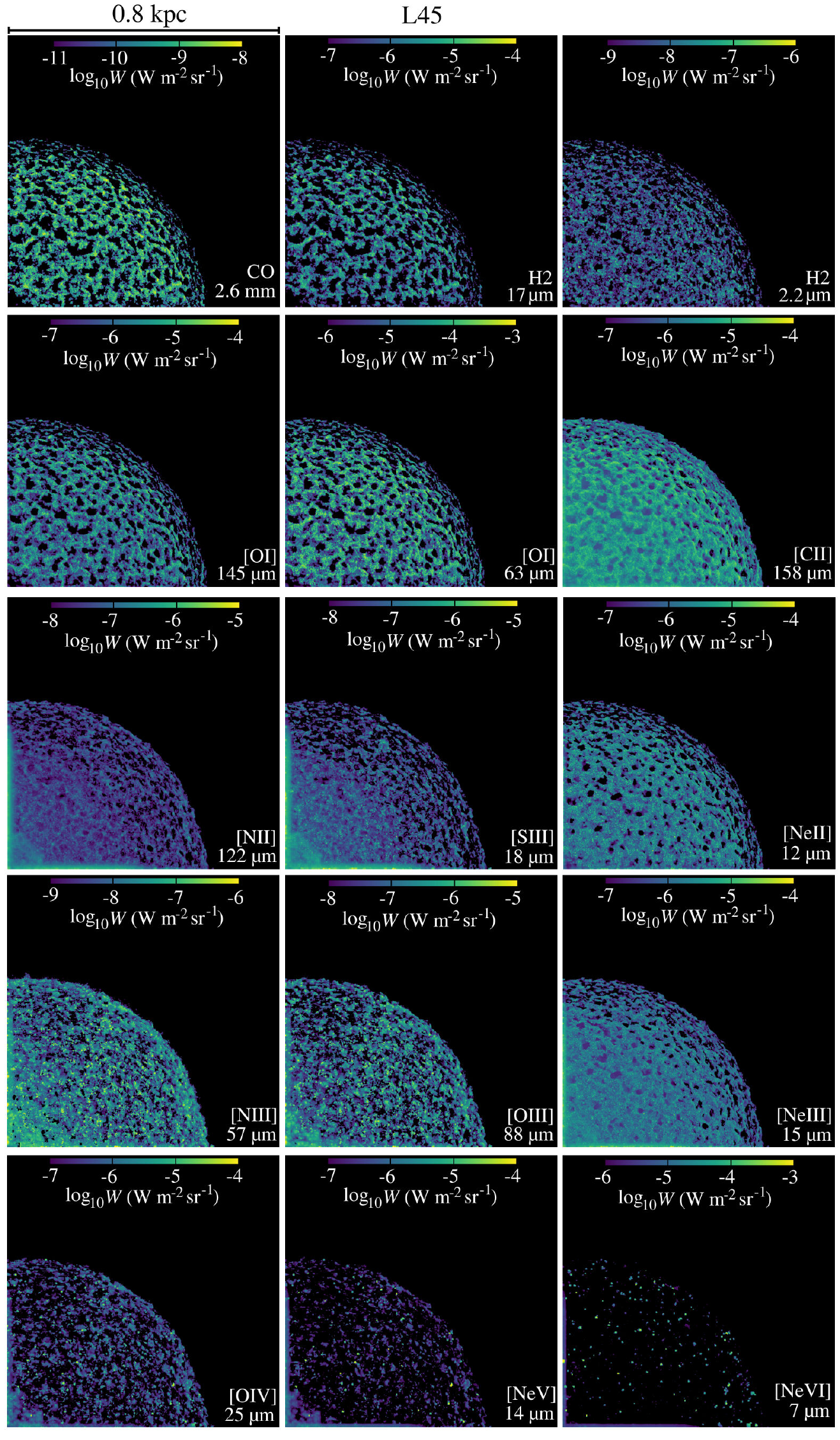}}
\mbox{
	\includegraphics[width=84mm]{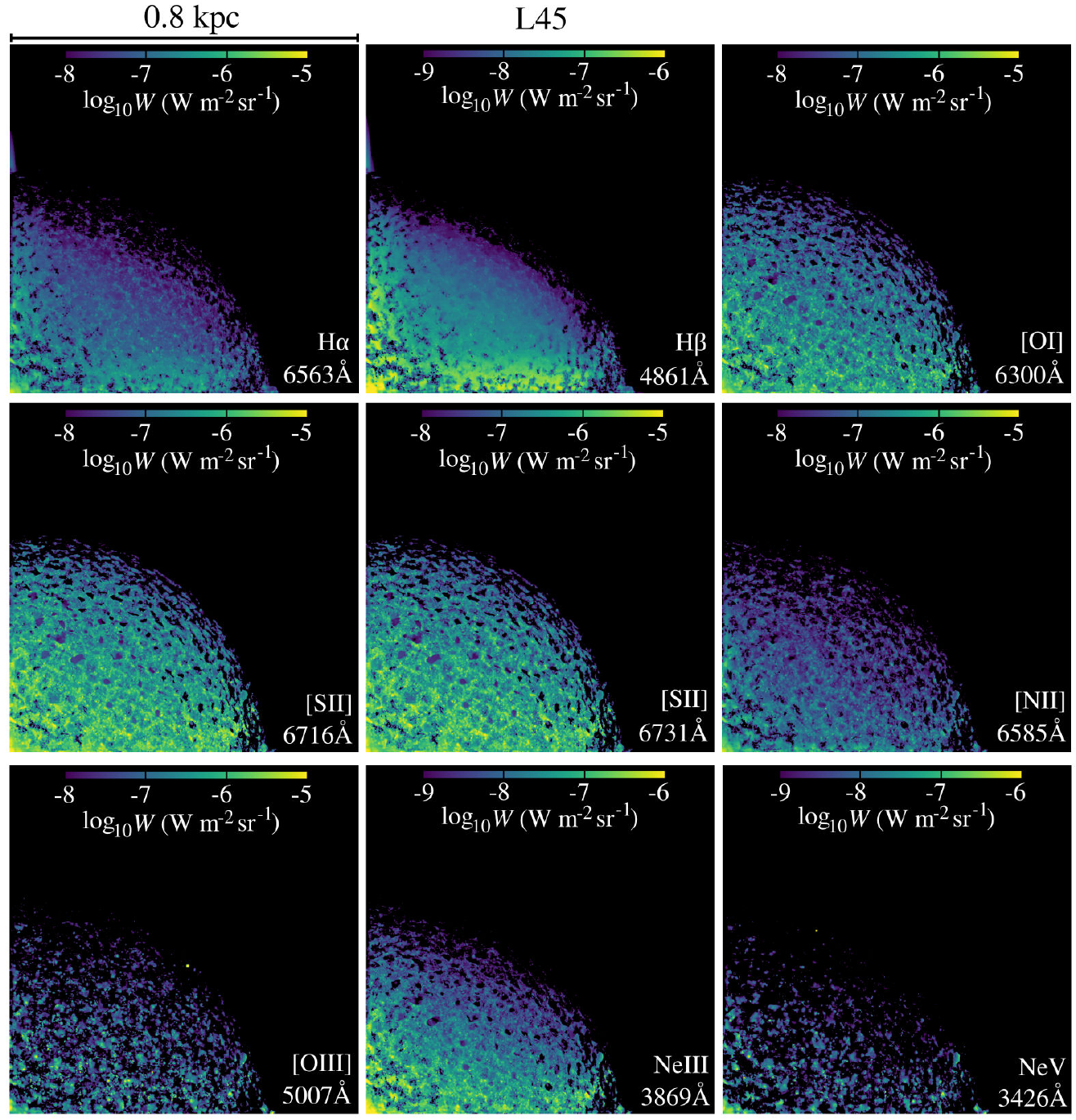}}
\vspace{-0.15 in}
\caption{Maps of the velocity-integrated continuum-subtracted emission from all millimetre, infrared, optical and UV emission lines in the full sample from simulation L45. 
\vspace{-0.15 in}} 
\label{full_images_L45}
\end{figure}

\begin{figure}
\centering
\mbox{
	\includegraphics[width=84mm]{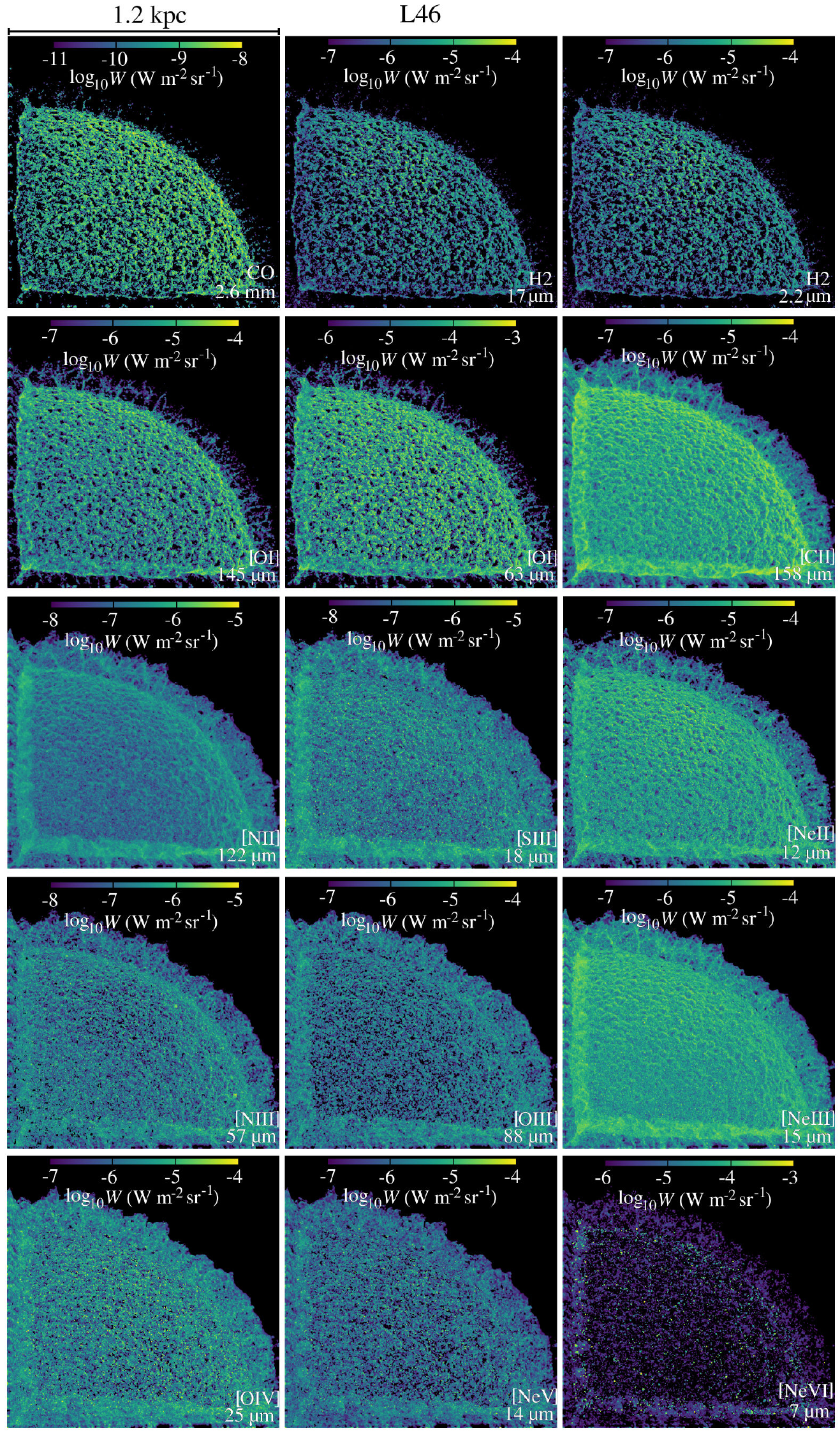}}
\mbox{
	\includegraphics[width=84mm]{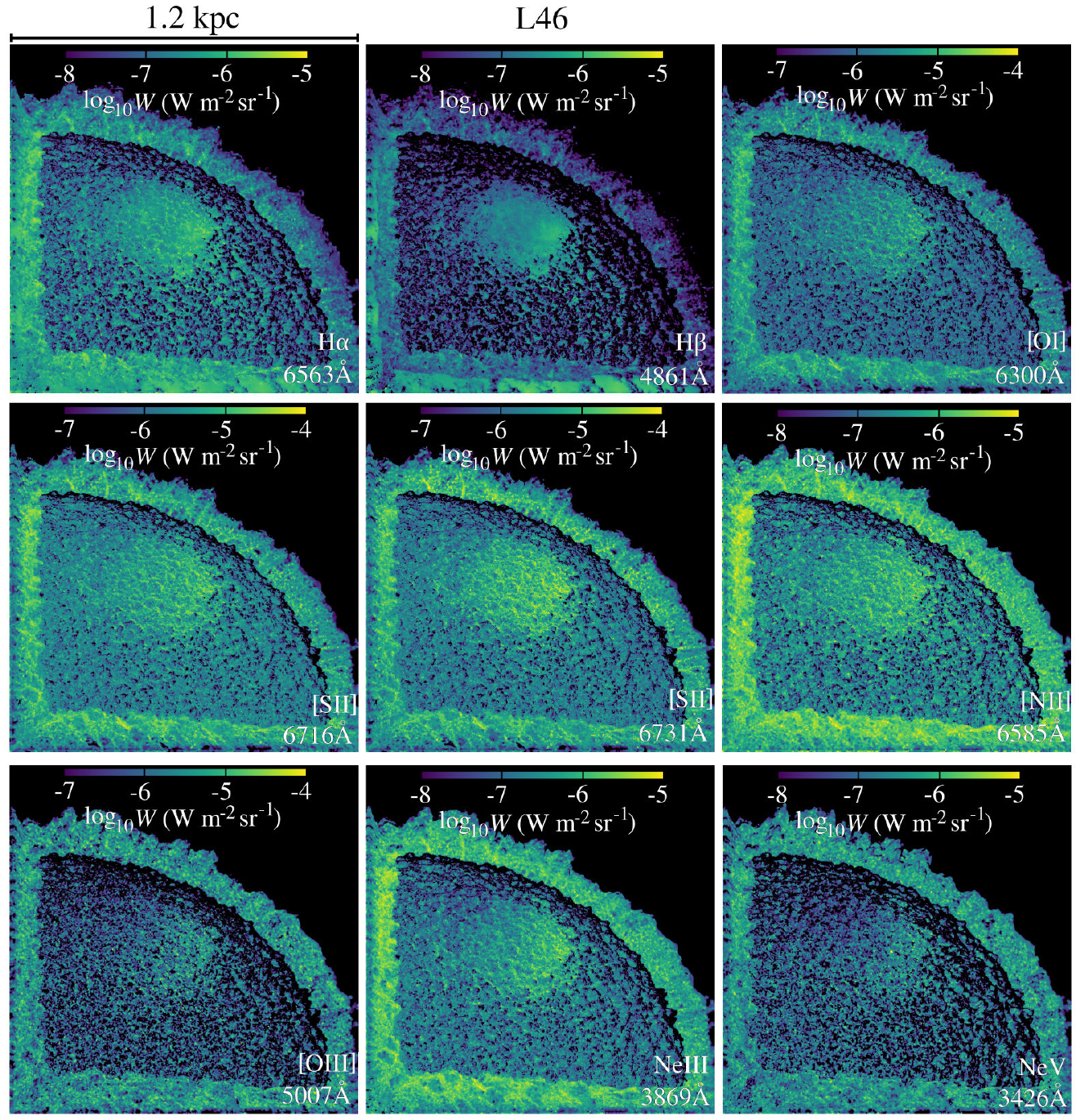}}
\vspace{-0.15 in}
\caption{As Fig.~\ref{full_images_L45}, but for the L46 simulation. 
\vspace{-0.15 in}} 
\label{full_images_L46}
\end{figure}

\begin{figure}
\centering
\mbox{
	\includegraphics[width=84mm]{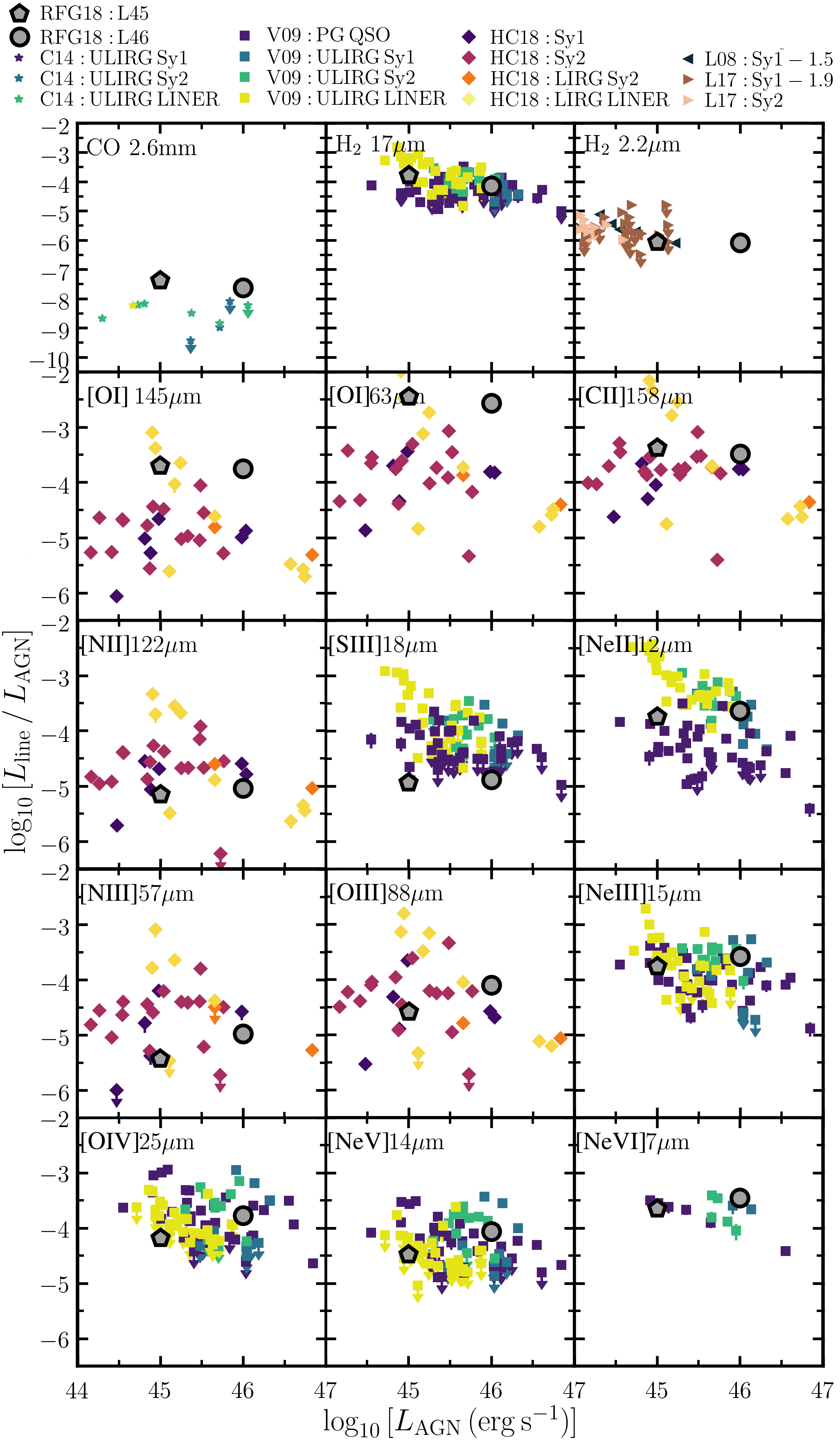}}
\vspace{-0.15 in}
\caption{Line luminosity ($L_{\rm{line}}$) divided by bolometric AGN luminosity ($L_{\rm{AGN}}$) versus $L_{\rm{AGN}}$ from the simulations (RFG18; grey symbols) and observed samples of AGN host galaxies from \citet{cicone14} (C14; stars), \citet{veilleux09} (V09; squares), \citet{herreracamus18} (HC18; diamonds), \citet{landt08} (L08; left triangles) and \citet{lamperti17} (L17; right triangles). The observations are divided according to AGN classification, as indicated by the colour. Upper limits are denoted by arrows. 
\vspace{-0.15 in}} 
\label{full_IR_vs_Lagn}
\end{figure}

\label{lastpage}

\end{document}